\documentclass[fleqn,a4paper]{CSML}
\pdfoutput=1

\usepackage{lastpage}
\newcommand{\dOi}{13(3:28)2017}
\lmcsheading{}{1--\pageref{LastPage}}{}{}%
{Nov.~30,~2016}{Sep.~19, 2017}{}

\usepackage[hidelinks]{hyperref}
\usepackage{amsmath}
\usepackage{amssymb}
\usepackage{array}
\usepackage[utf8]{inputenc}
\usepackage[T1]{fontenc}
\usepackage{bussproofs}
\usepackage{calc}
\usepackage{cite}
\usepackage{etex}
\usepackage{multirow}
\usepackage{graphicx}
\usepackage[all]{xy}
\usepackage{tikz}
\usetikzlibrary{positioning,shapes,automata,arrows,fit}
\usepackage{url}

\title{Formal Languages, Formally and Coinductively}
\author{Dmitriy Traytel}
\address{Institute of Information Security,
	Department of Computer Science,
	ETH Zürich, Switzerland}
\email{traytel@inf.ethz.ch}

\newbox\USCR
\newbox\USCRB
\let\oldunderscore=\_

\def\setlooseness#1{\looseness=#1 \immediate\write-1{@looseness=#1}}

\makeatletter
\newcommand{\Rmnum}[1]{\expandafter\@slowromancap\romannumeral #1@}
\makeatother

\EnableBpAbbreviations

\def\False{\bot}
\def\True{\top}

\newcommand\keyw[1]{\texttt{\def\_{{\footnotesize\oldunderscore}}#1}}
\newcommand\const[1]{\mathsf{\def\_{{\scriptstyle\oldunderscore}}#1}}
\newcommand\ty[1]{\mathit{\def\_{{\scriptstyle\oldunderscore}}#1}}

\newcommand{\omicron}{o}
\renewcommand{\emptyset}{\varnothing}
\renewcommand{\epsilon}{\varepsilon}
\renewcommand{\phi}{\varphi}

\newcommand\List[1]{[#1]}
\newcommand\Nil{\List{}}
\newcommand{\ConsOp}{\mathbin{\#}}
\newcommand{\AppendOp}{\mathbin{+\kern-4pt+}}
\newcommand{\Cons}[2]{{#1}
    {#2}}
\newcommand{\length}[1]{{|{#1}|}}
\newcommand{\map}{{\const{map}}}
\newcommand{\Inl}{{\const{Inl}}}
\newcommand{\Inr}{{\const{Inr}}}
\newcommand{\coiter}{{\const{coiter}}}
\newcommand{\corec}{{\const{corec}}}
\newcommand{\Id}{{\const{id}}}
\newcommand{\If}{{\keyw{if}}}
\newcommand{\Then}{{\keyw{then}}}
\newcommand{\Else}{{\keyw{else}}}
\newcommand{\Case}{{\keyw{case}}}
\newcommand{\Of}{{\keyw{of}}}
\newcommand{\Let}{{\keyw{let}}}

\newcommand{\AND}{{\keyw{and}}}
\newcommand{\Codatatype}{{\keyw{codatatype}}}
\newcommand{\Definition}{{\keyw{definition}}}
\newcommand{\Primrec}{{\keyw{primrec}}}
\newcommand{\Primcorec}{{\keyw{primcorec}}}
\newcommand{\Where}{{\keyw{where}}}
\newcommand{\Theorem}{{\keyw{theorem}}}
\newcommand{\Def}{{\keyw{def}}}

\newcommand{\Show}{{\keyw{show}}}
\newcommand{\Obtain}{{\keyw{obtain}}}
\newcommand{\By}{{\keyw{by}}}
\newcommand{\Fix}{{\keyw{fix}}}
\newcommand{\Assume}{{\keyw{assume}}}
\newcommand{\IsarProof}{{\keyw{proof}}}
\newcommand{\kQed}{{\keyw{qed}}}

\newcommand{\Simp}{{\mathit{simp}}}
\newcommand{\Auto}{{\mathit{auto}}}
\newcommand{\Rule}{{\mathit{rule}}}
\newcommand{\Coinduction}{{\mathit{coinduction}}}
\newcommand{\Arbitrary}{{\mathit{arbitrary}}}
\newcommand{\ArbitraryMod}{{\Arbitrary\!:}}
\newcommand{\RuleMod}{{\Rule\!:}}

\newcommand{\langCI}{{\mathit{coinduct}_\ty{lang}}}
\newcommand{\lessCI}{{\mathit{coinduct}_{\leq}}}
\newcommand{\langCIeq}{{\mathit{coinduct}^=_\ty{lang}}}
\newcommand{\langCIplus}{{\mathit{coinduct}^\PlusOp_\ty{lang}}}
\newcommand{\langCIreg}{{\mathit{coinduct}^{\bullet}_\ty{lang}}}
\newcommand{\lessCIreg}{{\mathit{coinduct}^{\bullet}_{\leq}}}

\newcommand{\LOR}{\mathrel\lor}
\newcommand{\LAND}{\mathrel\land}
\newcommand{\RIMP}{\longrightarrow}
\newcommand{\CUP}{\mathrel\cup}
\newcommand{\CAP}{\mathrel\cap}
\newcommand{\NOT}{\neg}
\newcommand{\CDOT}{\mathrel\cdot}
\newcommand{\IFF}{=}
\newcommand{\COLON}{\mathrel{:\mskip-.5mu:}}
\newcommand{\TAB}{\phantom{\Let}}
\newcommand{\MOD}{\mathrel{\const{mod}}}

\newcommand{\boolT}{{\ty{bool}}}
\newcommand{\listT}[1]{{{#1}\,\ty{list}}}
\newcommand{\setT}[1]{{{#1}\,\ty{set}}}
\newcommand{\langT}[1]{{{#1}\,\ty{lang}}}

\newcommand{\Zero}{\emptyset} 
\newcommand{\One}{\epsilon} 
\newcommand{\AtomOp}{{\const{A}}}
\newcommand{\Atomx}[1]{\AtomOp\;{#1}}
\newcommand{\Atom}[1]{{#1}}
\newcommand{\PlusOp}{{+}}
\newcommand{\Plus}[2]{{{#1}\mathrel{\PlusOp}{#2}}}
\newcommand{\Plusg}[2]{{{#1}\mathrel{\colorbox{lightgray}{\PlusOp}}{#2}}}
\newcommand{\TimesOp}{{\CDOT}}
\newcommand{\Times}[2]{{{#1}\mathrel{\TimesOp}{#2}}}
\newcommand{\Star}[1]{{{#1}^*}}
\newcommand{\StarOp}{{}^*}
\newcommand{\InterOp}{{\CAP}}
\newcommand{\Inter}[2]{{{#1}\mathrel{\InterOp}{#2}}}
\newcommand{\Not}[1]{{\overline{#1}}}
\newcommand{\NotOp}{\Not{\phantom{X}}}

\newcommand{\TimesLROp}{{\hat{\CDOT}}}
\newcommand{\TimesLR}[2]{{{#1}\mathrel{\TimesLROp}{#2}}}
\newcommand{\TimesPlusOp}{{\hat{\oplus}}}
\newcommand{\TimesPlus}[1]{{\TimesPlusOp\;{#1}}}
\newcommand{\StarLROp}{{\hat{*}}}
\newcommand{\StarLR}[2]{{{#1}\mathrel{\StarLROp}{#2}}}

\newcommand{\ShuffleOp}{{\parallel}}
\newcommand{\Shuffle}[2]{{{#1}\mathrel{\ShuffleOp}{#2}}}
\newcommand{\ShuffleLROp}{{\hat{\ShuffleOp}}}
\newcommand{\ShuffleLR}[2]{{{#1}\mathrel{\ShuffleLROp}{#2}}}

\newcommand{\reflCl}[1]{{#1}^{{}^=\kern-3pt}}
\newcommand{\PlusCl}[1]{{#1}^{{}^\PlusOp\kern-3pt}}
\newcommand{\PlusClsim}[1]{{#1}_{_\leq\kern-3pt}^{{}^\bullet\kern-3pt}}
\newcommand{\regCl}[1]{{#1}^{{}^\bullet\kern-3pt}}

\newcommand{\inLangOp}{{\in\nobreak\kern-1.15ex\in}}
\newcommand{\inLang}[2]{{#1}\mathrel{\inLangOp}{#2}}
\newcommand{\fromLangOp}{\const{out}}
\newcommand{\fromLang}[1]{\fromLangOp\;{#1}}
\newcommand{\toLangOp}{\const{in}}
\newcommand{\toLang}[1]{\toLangOp\;{#1}}

\newcommand{\finalOp}{{\omicron}}
\newcommand{\final}[1]{{\finalOp\;#1}}

\newcommand{\LangC}{{\const{L}}}
\newcommand{\derivOp}{{\delta}}
\newcommand{\derivx}[1]{\derivOp\;{#1}}
\newcommand{\deriv}[2]{\derivOp\;{#1}\;{#2}}

\newcommand{\vEven}{\mathit{even}}
\newcommand{\vOdd}{\mathit{odd}}

\newcommand{\vxs}{\mathit{w}}

\newcommand{\PARA}[1]{\paragraph*{\emph{#1}}}

\newenvironment{quotex}{\vskip\abovedisplayskip
    \begin{quote}}{\end{quote}\vskip\belowdisplayskip
}

\begin{document}

\begin{abstract}
Traditionally, formal languages are defined as sets of words. More recently, the
alternative coalgebraic or coinductive representation as infinite tries, i.e.,
prefix trees branching over the alphabet, has been used to obtain compact and
elegant proofs of classic results in language theory. In this article, we study
this representation in the Isabelle proof assistant. We define regular
operations on infinite tries and prove the axioms of Kleene algebra for those
operations. Thereby, we exercise corecursion and coinduction and confirm the
coinductive view being profitable in formalizations, as it improves over the
set-of-words view with respect to proof automation.
\end{abstract}

\maketitle

\section{Introduction}
\label{sec:intro}

If we ask a computer scientist what a formal language is, the answer will most
certainly be: a set of words. Here, we advocate another valid answer: an
infinite trie. This is the coalgebraic approach to
languages~\cite{Rutten98}, viewed through the lens of a lazy functional programmer.

This article shows how to formalize the coalgebraic or coinductive approach to
formal languages in the Isabelle/HOL proof assistant in the form of a gentle
introduction to corecursion and coinduction. Our interest in the coalgebraic
approach to formal languages arose in the context of a larger formalization effort of coalgebraic decision procedures for regular
languages~\cite{Traytel-CSL15,phd-traytel}. Indeed, we present here a reusable
library modeling languages, which lies at the core of those formalized decision
procedures. A lesson we have learned from this exercise and hope to convey here
is that often it is worthwhile to look at well-understood objects from a
different (in this case coinductive) perspective.

The literature is abound with tutorials on coinduction. So why bother writing
yet another one? First, because we finally can do it in Isabelle\slash HOL,
which became a coinduction-friendly proof assistant
recently~\cite{BlanchetteHLPPT-ITP14}. Earlier studies of coinduction in
Isabelle had to engage in tedious manual constructions just to define a
coinductive datatype~\cite{Paulson97}. Second, coinductive techniques are still not as
widespread as they could be (and we believe should be, since they constitute a
convenient proof tool for questions about semantics). Third, many
tutorials~\cite{Jacobs1997,KozenS12,Hinze10,Chlipala13,GimenezC98,setzer15},
with or without a theorem prover, exercise streams to the extent that one starts
to believe having seen every single stream example one can imagine. 
In contrast, Rutten~\cite{Rutten98} demonstrates that it is entirely feasible to
start a tutorial with a structure slightly more complicated than streams, but
familiar to every computer scientist. Moreover, Rot, Bonsangue,
and Rutten \cite{RotBR13,RotBR16} present an accessible introduction to
coinduction up to congruence using the coinductive view of formal languages.
Our work additionally focuses on
corecursion up-to and puts Rutten's exposition in the context of a proof
assistant.

When programming with infinite structures in the total setting of a proof
assistant, productivity must be ensured. Intuitively, a corecursive function is
productive if it always eventually yields observable output, e.g., in form of a
constructor. %
Functions that output exactly one constructor before proceeding corecursively or stopping with a fixed (non-corecursive) value
are called primitively corecursive---a fragment dual to well-understood
primitively recursive functions on inductive datatypes. Primitively corecursive
functions are productive. While sophisticated methods involving domain,
measure, and category theory for handling more complex corecursive
specifications have been proposed~\cite{HoelzlL-ITP14,BlanchettePT-ICFP15} and implemented in Isabelle~\cite{BlanchetteBLPT}, we
explore here how far primitive corecursion can get us. Restricting ourselves to
this fragment is beneficial in several ways. First, our constructions become
mostly Isabelle independent, since primitive corecursion is supported by all
coinduction-friendly proof assistants. Second, when working in the restricted
setting, we quickly hit and learn to understand the limits. In fact, we will face
some non-primitively corecursive specifications on infinite tries, which we
reduce to a composition of primitively corecursive specifications. Those
reductions are insightful and hint at a general pattern for handling certain
non-primitively corecursive specifications.

Infinite data structures are often characterized in terms of observations. For
infinite tries, which we define as a coinductive datatype or short codatatype
(Section~\ref{sec:3:trie}), we can observe the root, which in our case is
labeled by a Boolean value. This label determines if the empty word is
\emph{accepted} by the trie. Moreover, we can observe the immediate subtrees of
a trie, of which we have one for each alphabet letter. This observation
corresponds to making transitions in an automaton or rather computing the left
quotient $L_a = \{w \mid \Cons{a}{w} \in L\}$ of the language $L$ by the letter
$a$. Indeed, we will see that Brzozowski's ingenious derivative
operation~\cite{Brzozowski64}, which mimics this computation recursively on the
syntax of regular expressions, arises very naturally when defining regular
operations corecursively on tries (Section~\ref{sec:3:corec}). To validate our
definitions, we formally prove by coinduction that they satisfy the axioms of
Kleene algebra. Thereby, we use two similar but different coinduction principles for equalities and inequalities (Section~\ref{sec:3:coind}). %
After having presented our formalization, we step back and connect concrete
intuitive notions (such as tries) with abstract coalgebraic terminology
(Section~\ref{sec:3:coalg}). Furthermore, we discuss our formalization and its
relation to other work on corecursion and coinduction with or without proof
assistants~(Section~\ref{sec:3:discuss}). %

This article extends the homonymous FSCD 2016 paper~\cite{Traytel-FSCD16} with the definition of the shuffle product on tries (Subsection~\ref{subsec:shuffle}), a trie construction for context-free grammars (Subsection~\ref{subsec:cfg}), the coinductive treatment of inequalities (Subsection~\ref{ssec:3:sim}) instead of reducing them to equalities, and the proof that the trie construction for context-free grammars is sound~(Subsection~\ref{ssec:3:cfg}). 
The material presented here is based on the publicly available
Isabelle\slash HOL formalization~\cite{Traytel-AFP13} and is partly described in
the author's Ph.D.\ thesis~\cite{phd-traytel}.

\PARA{Preliminaries}

Isabelle/HOL is a proof assistant for higher-order logic, built around a small
trusted inference kernel. The kernel accepts only non-recursive type and
constant definitions. High-level specification mechanisms, which allow the user
to enter (co)recursive specifications, reduce this input to something equivalent
but non-recursive. The original (co)recursive specification is later derived as
a theorem. For a comprehensive introduction to Isabelle/HOL we refer to a recent
textbook~\cite[Part \Rmnum{1}]{Concrete}.

In Isabelle/HOL types $\tau$ are built from type variables $\alpha$, $\beta$,
etc., via type constructors~$\kappa$ written postfix (e.g., $\alpha\;\kappa$).
Some special types are the sum type $\alpha + \beta$, the product type $\alpha
\times \beta$, and the function type $\alpha \to \beta$, for which the type
constructors are written infix. Infix operators bind less tightly than the
postfix or prefix ones. Other important types are the type of Booleans $\boolT$
inhabited by exactly two different values $\True$ (truth) and $\False$ (falsity)
and the types $\listT{\alpha}$ and $\setT{\alpha}$ of lists and sets of elements
of type $\alpha$. %
For Boolean connectives and sets common mathematical notation is used. A special
constant is equality ${=} \COLON \alpha \to \alpha \to \boolT$, which is
polymorphic (it exists for any type, including the function type, on which it is
extensional, i.e., $(\forall x.\; f\;x=g\;x) \RIMP f = g$). %
The functions $\Inl$ and $\Inr$ are the standard embeddings of $+$.
Lists are constructed from $\Nil\COLON \listT{\alpha}$ and $\ConsOp \COLON
\alpha\to\listT{\alpha}\to\listT{\alpha}$; the latter written infix and often
omitted, i.e., we write $\Cons{a}{w}$ for $a \ConsOp w$. Likewise, list concatenation $\AppendOp$ is written infix and may be omitted. The notation
$\length{\vxs}$ stands for the length of the list $\vxs$, i.e., $\length{\Nil} =
0$ and $\length{\Cons{a}{\vxs}}=1+\length{\vxs}$.

\section{Languages as Infinite Tries}
\label{sec:3:trie}

We define the type of formal languages as a codatatype of infinite tries, that
is, (prefix) trees of infinite depth branching over the alphabet. We represent
the alphabet by the type parameter~$\alpha$. Each node in a trie carries a
Boolean label, which indicates whether the (finite) path to this node
constitutes a word inside or outside of the language. The function type models
branching: for each letter $x \COLON \alpha$ there is a subtree, which we call
\emph{$x$-subtree}. %
\begin{quotex}
$\Codatatype~\langT{\alpha} = \LangC~(\finalOp: \boolT)~(\derivOp: \alpha\to\langT{\alpha})$
\end{quotex}

The $\Codatatype$ command defines the type $\langT{\alpha}$ together with a
\emph{constructor} $\LangC\COLON\boolT \to (\alpha\to\langT{\alpha}) \to
\langT{\alpha}$ and two \emph{selectors} $\finalOp\COLON\langT{\alpha}\to\boolT$
and $\derivOp\COLON\langT{\alpha}\to\alpha\to\langT{\alpha}$. For a binary
alphabet $\alpha = \{a,\,b\}$, the trie $\vEven$ shown in
Figure~\ref{fig:3:even} is an inhabitant of $\langT{\alpha}$. The label
of its root is given by $\final{\vEven} = \True$ and its subtrees by another
trie $\deriv{\vEven}{a} = \deriv{\vEven}{b} = \vOdd$. Similarly, we have
$\final{\vOdd} = \False$ and $\deriv{\vOdd}{a} = \deriv{\vOdd}{b} = \vEven$.
Note that we could have equally written $\vEven =
\LangC\;\True\;(\lambda\_.\;\vOdd)$ and $\vOdd =
\LangC\;\False\;(\lambda\_.\;\vEven)$ to obtain the same mutual characterization
of $\vEven$ and $\vOdd$.

\begin{figure}[h]
\centering
\begin{tikzpicture}[node distance=3em]
\node (eps) {$\True$};
\node[below left of=eps, xshift=-2em] (a) {$\False$};
\node[below right of=eps, xshift=2em] (b) {$\False$};
\node[below left of=a] (aa) {$\True$};
\node[below right of=a] (ab) {$\True$};
\node[below left of=b] (ba) {$\True$};
\node[below right of=b] (bb) {$\True$};
\node[below of=aa, yshift=2ex] (aax) {$\ldots$};
\node[below of=ab, yshift=2ex] (abx) {$\ldots$};
\node[below of=ba, yshift=2ex] (bax) {$\ldots$};
\node[below of=bb, yshift=2ex] (bbx) {$\ldots$};

\path[-latex,draw] (eps) -- node[above left,gray] {$a\vphantom{b}$} (a);
\path[-latex,draw] (eps) -- node[above right,gray] {$b$} (b);
\path[-latex,draw] (a) -- node[above left,gray] (lab) {$a\vphantom{b}$} (aa);
\path[-latex,draw] (a) -- node[above right,gray] {$b$} (ab);
\path[-latex,draw] (b) -- node[above left,gray] {$a\vphantom{b}$} (ba);
\path[-latex,draw] (b) -- node[above right,gray] {$b$} (bb);
\path[-latex,draw] (aa) -- node[yshift=.5ex,left,gray] {$a\vphantom{b}$} (aax.west);
\path[-latex,draw] (aa) -- node[yshift=.5ex,right,gray] {$b$} (aax.east);
\path[-latex,draw] (ab) -- node[yshift=.5ex,left,gray] {$a\vphantom{b}$} (abx.west);
\path[-latex,draw] (ab) -- node[yshift=.5ex,right,gray] {$b$} (abx.east);
\path[-latex,draw] (ba) -- node[yshift=.5ex,left,gray] {$a\vphantom{b}$} (bax.west);
\path[-latex,draw] (ba) -- node[yshift=.5ex,right,gray] {$b$} (bax.east);
\path[-latex,draw] (bb) -- node[yshift=.5ex,left,gray] {$a\vphantom{b}$} (bbx.west);
\path[-latex,draw] (bb) -- node[yshift=.5ex,right,gray] {$b$} (bbx.east);
\end{tikzpicture}
\caption{Infinite trie $\vEven$}
\label{fig:3:even}
\end{figure}

We gave our type the name $\langT{\alpha}$, to remind us to think of its
inhabitants as formal languages. In the following, we use the terms language and
trie synonymously.

Beyond defining the type and the constants, the $\Codatatype$ command also
exports a wealth of properties about them such as $\final{(\LangC\;b\;d)} = b$,
the injectivity of $\LangC$, or more interestingly the coinduction rule.
Informally, coinduction allows us to prove equality of tries which cannot be
distinguished by finitely many selector applications.

Clearly, we would like to identify the trie $\vEven$ with the regular language
of all words of even length $\{w \in \{a,\,b\}^* \mid \length{w} \MOD 2 = 0\}$,
also represented by the regular expression
$\Star{(\Times{(\Plus{\Atom{a}}{\Atom{b}})}{(\Plus{\Atom{a}}{\Atom{b}})})}$.
Therefore, we define the notion of word membership $\inLangOp$ on tries by
primitive (or structural) recursion on the word using Isabelle's $\Primrec$ command. %
\begin{quotex}
$\Primrec~\inLangOp {}\COLON \listT{\alpha}\to\langT{\alpha}\to\boolT~\Where\\
\TAB \inLang{\Nil}{L} = \final{L}\\
\TAB \inLang{\Cons{a}{w}}{L} = \inLang{w}{\deriv{L}{a}}$
\end{quotex}

Using $\inLangOp$, each trie can be assigned a language in the traditional set
of lists view.
\begin{quotex}
$\Definition~\fromLangOp \COLON~ \langT{\alpha} \to \setT{\listT{\alpha}}~\Where\\
\TAB \fromLang{L} = \{w \mid \inLang{w}{L}\}$
\end{quotex}
With this definition, we obtain $\fromLang{\vEven} = \{w \in \{a,\,b\}^* \mid
\length{w} \MOD 2 = 0\}$.


\section{Operations on Tries}
\label{sec:3:corec}

So far, we have only specified some concrete infinite tries informally.
Formally, we will use primitive corecursion, which is dual to primitive
recursion. Primitively recursive functions consume one constructor before
proceeding recursively. Primitively corecursive functions produce one
\emph{guarding} constructor whose arguments are allowed to be either
non-recursive terms or a corecursive call (applied to arbitrary non-recursive
arguments). We call a function truly primitively corecursive if not all of the constructor's argument are non-recursive. %
The $\Primcorec$ command reduces a primitively corecursive specification to a
non-recursive definition, which is accepted by Isabelle's inference
kernel~\cite{BlanchetteHLPPT-ITP14}. Internally, the reduction employs a
dedicated combinator for primitive corecursion on tries generated by the
$\Codatatype$ command. The $\Primcorec$ command slightly relaxes the above  restriction of primitive corecursion by allowing syntactic
conveniences, such as lambda abstractions, $\Case$-, and $\If$-expressions, to
appear between the guarding constructor and the corecursive call. The relaxation is resembling the syntactic guardedness check used in type theory~\cite[Section~2.3]{Coquand93}, however still allowing only exactly one constructor to guard a corecursive call.

\subsection{Primitively Corecursive Operations}

We start with some simple examples: the languages of the base cases of regular
expressions. Intuitively, the trie $\Zero$ representing the empty language is
labeled with $\False$ everywhere and the trie $\One$ representing the empty word
language is labeled with $\True$ at its root and with $\False$ everywhere else.
The trie $\Atomx{a}$ representing the singleton language of the one letter word
$a$ is labeled with $\False$ everywhere except for the root of its $a$-subtree.
This intuition is easy to capture formally. %
\begin{quotex}
\begin{tabular}{@{}>{\raggedright\arraybackslash}p{0.34\textwidth}>{\raggedright\arraybackslash}p{0.34\textwidth}@{}}
$\Primcorec~\Zero {}\COLON \langT{\alpha}~\Where\allowbreak
\TAB \Zero = \LangC~\False~(\lambda x.\;\Zero)$&
$\Primcorec~\One {}\COLON \langT{\alpha}~\Where\allowbreak
\TAB \One = \LangC~\True~(\lambda x.\;\Zero)$
\end{tabular}\\[2\jot]
$\Primcorec~\AtomOp {}\COLON \alpha \to \langT{\alpha}~\Where\\
\TAB \final{(\Atomx{a})} = \False\\
\TAB \derivx{(\Atomx{a})} = \lambda x.\;\If\;a = x\;\Then\;\One\;\Else\;\Zero$
\end{quotex}
Among these three definitions only $\Zero$ is truly primitively corecursive.

The specifications of $\Zero$ and $\One$ differ syntactically from the one of $\AtomOp$.
The constants $\Zero$ and $\One$ are defined using the so called
\emph{constructor view}. The constructor view allows the user to enter equations of the form
constant or function equals constructor, where the arguments of the constructor
are restricted as described above. Such
definitions should be familiar to any (lazy) functional programmer.

In contrast, the specification of $\AtomOp$ is expressed in the \emph{destructor
view}. Here, we specify the constant or function being defined by observations
or experiments via \emph{selector equations}. The allowed experiments on a trie
are given by its selectors $\finalOp$ and $\derivOp$. We can observe the label
at the root and the subtrees. Specifying the observation for each
selector---again restricted either to be a non-recursive term or to contain the
corecursive call only at the outermost position (ignoring lambda abstractions, $\Case$-, and $\If$-expressions)---yields a unique
characterization of the function being defined.

It is straightforward to rewrite specifications in either of the views into the
other one. The $\Primcorec$ command performs this rewriting internally and
outputs the theorems corresponding to the user's input specification in both
views. The constructor view theorems serve as executable code equations.
Isabelle's code generator~\cite{HaftmannN10} can use these equations to generate
code which make sense in programming languages with lazy evaluation. In
contrast, the destructor view offers safe simplification rules even when applied
eagerly during rewriting as done by Isabelle's simplifier. Note that constructor
view specifications such as $\Zero = \LangC~\False~(\lambda x.\;\Zero)$ will
cause the simplifier to loop when applied eagerly.

Now that the basic building blocks $\Zero$, $\One$, and $\AtomOp$ are in place,
we turn our attention to how to combine them to obtain more complex languages.
We start with the simpler combinators for union, intersection, and complement,
before moving to the more interesting concatenation and iteration. The union
$\PlusOp$ of two tries should denote set union of languages (i.e.,
$\fromLang{(\Plus{L}{K})} = \fromLang{L} \CUP \fromLang{K}$ should hold). It is
defined corecursively by traversing the two tries in parallel and computing for
each pair of labels their disjunction. Intersection $\InterOp$ is analogous.
Complement $\NotOp$ simply inverts every label. 
\begin{quotex}
$\Primcorec~\PlusOp {}\COLON \langT{\alpha}\to\langT{\alpha}\to\langT{\alpha}~\Where\\
\TAB \final{(\Plus{L}{K})} = \final{L} \LOR \final{K}\\
\TAB \derivx{(\Plus{L}{K})} = \lambda x.\;\Plus{\deriv{L}{x}}{\deriv{K}{x}}\\[2\jot]
\Primcorec~\InterOp {}\COLON \langT{\alpha}\to\langT{\alpha}\to\langT{\alpha}~\Where\\
\TAB \final{(\Inter{L}{K})} = \final{L} \LAND \final{K}\\
\TAB \derivx{(\Inter{L}{K})} = \lambda x.\;\Inter{\deriv{L}{x}}{\deriv{K}{x}}\\[2\jot]
\Primcorec~\NotOp {}\COLON \langT{\alpha}\to\langT{\alpha}~\Where\\
\TAB \final{\Not{L}} = \NOT\;\final{L}\\
\TAB \derivx{\Not{L}} = \lambda x.\;\Not{\deriv{L}{x}}$
\end{quotex}

Let us look at the specifying selector equations which we have seen so far from
a different perspective. Imagine $L$ and $K$ being not tries but instead
syntactic regular expressions, $\AtomOp$, $\PlusOp$, $\InterOp$, and $\NotOp$
constructors of a datatype for regular expressions, and $\finalOp$ and
$\derivOp$ two operations that we define recursively on this syntax. From that
perspective, the operations are familiar: rediscovered Brzozowski derivatives of
regular expressions~\cite{Brzozowski64} and the empty word acceptance (often
also called \emph{nullability}) test on regular expressions in the destructor view
equations for the selectors $\derivOp$ and $\finalOp$. 
There is an important difference, though: while Brzozowski derivatives work with
syntactic objects, our tries are semantic objects on which equality denotes
language equivalence. For example, we will later prove $\Plus{\Zero}{L} = L$ for
tries, whereas $\Plus{\Zero}{L} \not= L$ holds for regular expressions. The
coinductive view reveals that derivatives and the acceptance test are the two
fundamental ingredients that completely characterize regular languages and arise
naturally when only considering the semantics.

\subsection{Reducing Corecursion Up-to to Primitive Corecursion}
\label{subsec:upto}

Concatenation $\TimesOp$ is the next regular operation that we want
to define on tries. Thinking of Brzozowski derivatives and the acceptance test,
we would usually specify it by the following two equations.
\newcommand{\eqconc}{\eqref{eqconc}}
%
\begin{align}
\label{eqconc}
  \begin{split}
    \final{(\Times{L}{K})} &= \final{L} \LAND \final{K} \\
    \derivx{(\Times{L}{K})} &= \lambda x.\;\Plus{(\Times{\deriv{L}{x}}{K})}{
      (\If\;\final{L}\;\Then\;\deriv{K}{x}\;\Else\;\Zero)}
  \end{split}
\end{align}

A difficulty arises here, since this specification is not primitively
corecursive---the right hand side of the second equation contains a corecursive
call but not at the topmost position (but rather under $\PlusOp$ here). We call this
kind of corecursion \emph{up to $\PlusOp$}.

Without tool support for corecursion up-to, concatenation must be defined
dif\-fer\-ent\-ly---as a composition of other primitively corecursive
operations. Intuitively, we would like to separate the above $\derivOp$-equation
into two along the $\PlusOp$ and sum them up afterwards. Technically, the
situation is more involved. Since the $\derivOp$-equation is corecursive, we
cannot just create two tries by primitive corecursion.

Figure~\ref{fig:3:times} depicts the trie that should result from concatenating
an arbitrary trie $K$ to the concrete given trie $L$. Procedurally, the
concatenation must replace every subtree $T$ of $L$ that has $\True$ at the
root (those are positions where words from $L$ end) by the trie $\Plus{U}{K}$
where $U$ is the trie obtained from $T$ by changing its root from $\True$ to
$\final{K}$. For uniformity with the above $\derivOp$-equation, we imagine
subtrees $F$ of $L$ with label $\False$ at the root as also being replaced by
$\Plus{F}{\Zero}$, which, as we will prove later, has the same effect as leaving
$F$ alone.

\newcommand{\TreePic}[1]{
\raisebox{-1em}[0pt][0pt]{
\begin{tikzpicture}[scale=0.8]
\node at (.5,-0.25) {$#1$};
\node at (.5,-.55) {$\ldots$};
\path[draw, thick] (0,-.5) -- (.5,.5) -- (1,-.5);
\end{tikzpicture}
}}

\newcommand{\TreePicW}[1]{
\raisebox{-2em}[0pt][0pt]{
\begin{tikzpicture}[scale=0.8]
\node at (.5,-0.25) {$#1$};
\node at (.5,-.55) {$\ldots$};
\path[draw, thick] (-.5,-.5) -- (.5,.5) -- (1.5,-.5);
\end{tikzpicture}
}}

\begin{figure}
\centering
\begin{tikzpicture}[node distance=3em]
\node (eps) {$\True$};
\node[below left of=eps, xshift=-2em] (a) {$\True$};
\node[below right of=eps, xshift=2em] (b) {$\False$};
\node[below of=a, yshift=2ex] (ax) {$\ldots$};
\node[below of=b, yshift=2ex] (bx) {$\ldots$};

\path[-latex,draw] (eps) -- node[above left,gray] {$a\vphantom{b}$} (a);
\path[-latex,draw] (eps) -- node[above right,gray] {$b$} (b);
\path[-latex,draw] (a) -- node[yshift=.5ex,left,gray] {$a\vphantom{b}$} (ax.west);
\path[-latex,draw] (a) -- node[yshift=.5ex,right,gray] {$b$} (ax.east);
\path[-latex,draw] (b) -- node[yshift=.5ex,left,gray] {$a\vphantom{b}$} (bx.west);
\path[-latex,draw] (b) -- node[yshift=.5ex,right,gray] {$b$} (bx.east);
\end{tikzpicture}
\hspace*{7em}
\begin{tikzpicture}[node distance=3em]
\node (eps) {\rlap{$\Plus{\final{K}\;\quad}{\!\!\TreePic{K}}$}\phantom{$\final{K}$}};
\node[below left of=eps, xshift=-2em] (a) {\phantom{$\final{K}$}\llap{$\Plus{\TreePic{K}\!\!}{\;\quad\final{K}}$}};
\node[below right of=eps, xshift=2em] (b) {\rlap{$\Plus{\False\;\quad}{\!\!\TreePic{\Zero}}$}\phantom{$\False$}};
\node[below right of=eps, xshift=2em] (phantomb) {
\phantom{$\Plus{\TreePic{\Zero}\!\!}{\Plus{\quad\;\False\;\quad}{\!\!\TreePic{\Zero}}}$}};
\node[below of=a, yshift=2ex] (ax) {$\ldots$};
\node[below of=b, yshift=2ex] (bx) {$\ldots$};

\path[-latex,draw] (eps) -- node[above left,gray] {$a\vphantom{b}$} (a);
\path[-latex,draw] (eps) -- node[below left,gray] {$b$} (b);
\path[-latex,draw] (a) -- node[yshift=.5ex,left,gray] (lab) {$a\vphantom{b}$} (ax.west);
\path[-latex,draw] (a) -- node[yshift=.5ex,right,gray] {$b$} (ax.east);
\path[-latex,draw] (b) -- node[yshift=.5ex,left,gray] {$a\vphantom{b}$} (bx.west);
\path[-latex,draw] (b) -- node[yshift=.5ex,right,gray] {$b$} (bx.east);
\end{tikzpicture}
\caption{Tries for $L$ (left) and the concatenation $\Times{L}{K}$ (right)}
\label{fig:3:times}
\end{figure}

\begin{figure}
\centering
\begin{tikzpicture}[node distance=4em]
\node (eps) {$\final{K}$};
\node[below left of=eps, xshift=-2em] (a) {$\final{K}$};
\node[below left of=eps, xshift=-8em] (a') {$\TreePicW{\deriv{K}{a}}$};
\node[below right of=eps, xshift=2em] (b) {$\False$};
\node[below right of=eps, xshift=8em] (b') {$\TreePicW{\deriv{K}{b}}$};
\node[below of=a, yshift=2ex] (ax) {$\ldots$};
\node[below of=b, yshift=2ex] (bx) {$\ldots$};

\path[-latex,draw] (eps) -- node[yshift=-.5ex,right,gray] {$a\vphantom{b}$} (a);
\path[-latex,draw] (eps) -- node[yshift=-.5ex,left,gray] {$b$} (b);
\path[-latex,draw] (eps) -- node[above left,gray] {$a'\vphantom{b'}$} ([xshift=1ex,yshift=.5ex]a'.north);
\path[-latex,draw] (eps) -- node[above right,gray] {$b'$} ([xshift=-1ex,yshift=.5ex]b'.north);
\path[-latex,draw] (a) -- node[left,gray] {$a\vphantom{b}$} (ax.west);
\path[-latex,draw] (a) -- node[right,gray] {$b$} (ax.east);
\path[-latex,draw] (a) -- node[above,gray] {$a'\vphantom{b'}$} ([xshift=-1.4em,yshift=.7ex]ax.north west);
\path[-latex,draw] (a) -- node[above,gray,xshift=0.1em] {$b'$} ([xshift=1.4em,yshift=.7ex]ax.north east);
\path[-latex,draw] (b) -- node[left,gray] {$a\vphantom{b}$} (bx.west);
\path[-latex,draw] (b) -- node[right,gray] {$b$} (bx.east);
\path[-latex,draw] (b) -- node[above,gray] {$a'\vphantom{b'}$} ([xshift=-1.4em,yshift=.7ex]bx.north west);
\path[-latex,draw] (b) -- node[above,gray,xshift=0.1em] {$b'$} ([xshift=1.4em,yshift=.7ex]bx.north east);

\node[yshift=-1.5ex] at ([xshift=-1.5em]ax.north west) {$\TreePicW{\deriv{K}{a}}$};
\node[yshift=-1.5ex] at ([xshift=1.5em]ax.north east) {$\TreePicW{\deriv{K}{b}}$};
\node[yshift=-1em-1.5ex, xshift=.1ex] at ([xshift=-1.5em]bx.north west) {$\TreePic{\Zero}$};
\node[yshift=-1em-1.5ex, xshift=-.1ex] at ([xshift=1.5em]bx.north east) {$\TreePic{\Zero}$};

\node[below of=ax, yshift=2.7ex] {};
\end{tikzpicture}
\caption{Trie for deferred concatenation $\TimesLR{L}{K}$}
\label{fig:3:timesLR}
\end{figure}

{
\looseness=-1
Figure~\ref{fig:3:timesLR} presents one way to bypass the restrictions imposed
by primitive corecursion. We are not allowed to use $\PlusOp$ \emph{after} proceeding
corecursively, but we may arrange the arguments of $\PlusOp$ in a broader trie
over a doubled alphabet formally modeled by pairing letters of the alphabet with
a Boolean flag. In Figure~\ref{fig:3:timesLR} we write $a$ for
$(a,\,\True)$ and $a'$ for $(a,\,\False)$. Because it defers the summation, we
call this primitively corecursive procedure \emph{deferred concatenation}
$\TimesLROp$. %
\begin{quotex}
$\Primcorec~\TimesLROp {}\COLON \langT{\alpha}\to\langT{\alpha}\to\langT{(\alpha\times\boolT)}~\Where\\
\TAB \final{(\TimesLR{L}{K})} = \final{L} \LAND \final{K}\\
\TAB \derivx{(\TimesLR{L}{K})} = \lambda (x,\,b).\;\If\;b\;\Then\;\TimesLR{\deriv{L}{x}}{K}\;\Else\;
  \If\;\final{L}\;\Then\;\TimesLR{\deriv{K}{x}}{\One}\;\Else\;\Zero$\kern-2em
\end{quotex}
Note that unlike in the Figure~\ref{fig:3:timesLR}, where we informally plug the
trie $\deriv{K}{x}$ as some $x'$-subtrees, the formal definition must be more
careful with the types. More precisely, $\deriv{K}{x}$ is of type
$\langT{\alpha}$, while something of type $\langT{(\alpha\times\boolT)}$ is
expected. This type mismatch is resolved by further concatenating $\One$ to
$\deriv{K}{x}$ (again in a deferred fashion) without corrupting the intended
semantics.

}

Once the trie for the deferred concatenation has been built, the desired trie
for concatenation can be obtained by a second primitively corecursive traversal
that sums the $x$- and $x'$-subtrees \emph{before} proceeding corecursively. %
\begin{quotex}
$\Primcorec~\TimesPlusOp {}\COLON \langT{(\alpha\times\boolT)}\to\langT{\alpha}~\Where\\
\TAB \final{(\TimesPlus{L})} = \final{L}\\
\TAB \derivx{(\TimesPlus{L})} = \lambda x.\;\TimesPlus{(\Plus{\deriv{L}{(x,\,\True)}}{\deriv{L}{(x,\,\False)}})}$
\end{quotex}

Finally, we can define the concatenation as the composition of $\TimesLROp$ and
$\TimesPlusOp$. The earlier natural selector equations \eqconc{} for $\TimesOp$ are provable for this
definition. %
\begin{quotex}
$\Definition~\TimesOp {}\COLON \langT{\alpha}\to\langT{\alpha}\to\langT{\alpha}~\Where\\
\TAB \Times{L}{K} = \TimesPlus{(\TimesLR{L}{K})}$
\end{quotex}

The situation with iteration is similar. The selector equations following the
Brzozowski derivative of $\Star{L}$ yield a non-primitively corecursive
specification: it is corecursive up to $\TimesOp$.%
\newcommand{\eqstar}{\eqref{eqstar}}
%
\begin{align}
\label{eqstar}
\begin{split}
\final{(\Star{L})} &= \True\\
\derivx{(\Star{L})} &= \lambda x.\;\Times{\deriv{L}{x}}{\Star{L}}
\end{split}
\end{align}

As before, the restriction is circumvented by altering the operation slightly.
We define the binary operation \emph{deferred iteration} $\StarLR{L}{K}$, whose
language should represent $\Times{L}{\Star{K}}$ (although we have not defined
$\StarOp$ yet). When constructing the subtrees of $\StarLR{L}{K}$ we keep
pulling copies of the second argument into the first argument before proceeding
corecursively (the second argument itself stays unchanged). %
\begin{quotex}
$\Primcorec~\StarLROp {}\COLON \langT{\alpha}\to\langT{\alpha}\to\langT{\alpha}~\Where\\
\TAB \final{(\StarLR{L}{K})} = \final{L}\\
\TAB \derivx{(\StarLR{L}{K})} =
  \lambda x.\;\StarLR{(\deriv{(\Times{L}{(\Plus{\One}{K})})}{x})}{K}$
\end{quotex}

Supplying $\One$ as the first argument to $\StarLROp$ defines iteration
for which the original selector equations~\eqstar{} hold. %
\begin{quotex}
$\Definition~\_\StarOp {}\COLON \langT{\alpha}\to\langT{\alpha}~\Where\\
\TAB \Star{L} = \StarLR{\One}{L}$
\end{quotex}

We have defined all the standard regular operations on tries. Later we will
prove that those definitions satisfy the axioms of Kleene algebra, meaning that
they behave as expected. Already now we can compose the operations to define new tries, for example the introductory $\vEven = \Star{(\Times{(\Plus{\Atomx{a}}{\Atomx{b}})}{(\Plus{\Atomx{a}}{\Atomx{b}})})}$.

\subsection{Adding Further Operations}
\label{subsec:shuffle}

In the coinductive representation, adding new operations corresponds to
defining a new corecursive function on tries. Compared with adding a new
constructor to the inductive datatype of regular expressions and extending all
previously defined recursive functions on regular expressions to
account for this new case, this a is rather low-cost library extension.
Wadler called this tension between extending syntactic and semantic objects the
Expression Problem~\cite{wadler-expression}.\footnote{Note that codatatypes alone are not
the solution to the Expression Problem---they just populate the other side of
the spectrum with respect to datatypes. In fact, adding new selectors to
tries would be as painful as adding new constructors to the datatype of regular
expressions. %
Rendel et al.~\cite{RendelTO15} outline how automatic conversions between the
inductive and the coinductive view can help solving the Expression Problem.}

As an example library extension, we define the regular shuffle product operation on
languages, which adheres to the following selector equations.
\newcommand{\eqshuffle}{\eqref{eqshuffle}}
%
\begin{align}
\label{eqshuffle}
\begin{split}
\final{(\Shuffle{L}{K})} &= \final{L} \LAND \final{K} \\
\derivx{(\Shuffle{L}{K})} &= \lambda x.\;\Plus{(\Shuffle{\deriv{L}{x}}{K})}{
    (\Shuffle{L}{\deriv{K}{x}})}
\end{split}
\end{align}
Intuitively, for words $\inLang{w}{L}$ and $\inLang{v}{K}$, the shuffle product $\Shuffle{L}{K}$ contains all possible interleavings of $w$ and $v$.
As it is the case for concatenation, the selector equations for the shuffle product $\ShuffleOp$ are corecursive up to $\PlusOp$. Thus, as for $\TimesOp$, we first define a \emph{deferred shuffle product} operation, which keeps the $\PlusOp$ occurring outside of the corecursive calls in the second equation ``unevaluated'' by using the doubled alphabet $\alpha \times \boolT$ instead of $\alpha$.
\begin{quotex}
    $\Primcorec~\ShuffleLROp {}\COLON \langT{\alpha}\to\langT{\alpha}\to\langT{(\alpha\times\boolT)}~\Where\\
    \TAB \final{(\ShuffleLR{L}{K})} = \final{L} \LAND \final{K}\\
    \TAB \derivx{(\ShuffleLR{L}{K})} = \lambda (x,\,b).\;\If\;b\;\Then\;\ShuffleLR{\deriv{L}{x}}{K}\;\Else\;\ShuffleLR{L}{\deriv{K}{x}}$
\end{quotex}
{
\looseness=-1
A second primitively corecursive traversal sums the $(a,\,\True)$- and $(a,\,\False)$-subtrees using the same function $\TimesPlusOp$ as in the definition of concatenation. Then, the shuffle product can be defined as the composition of $\ShuffleLROp$ and $\TimesPlusOp$.

}
\begin{quotex}
    $\Definition~\ShuffleOp {}\COLON \langT{\alpha}\to\langT{\alpha}\to\langT{\alpha}~\Where\\
    \TAB \Shuffle{L}{K} = \TimesPlus{(\ShuffleLR{L}{K})}$
\end{quotex}

\subsection{Context-Free Grammars as Tries}
\label{subsec:cfg}

\newcommand{\CFG}{\mathsf{G}}
\newcommand{\CFGTT}{\mathcal{T}}
\newcommand{\CFGNT}{\mathcal{N}}
\newcommand{\CFGNs}{\mathit{N}}
\newcommand{\CFGS}{\mathsf{S}}
\newcommand{\CFGP}{\mathsf{P}}
\newcommand{\CFGOp}{\mathsf{mk}}
\newcommand{\CFGxOp}{\mathsf{mk0}}
\newcommand{\CFGx}[1]{\mathsf{x}(#1)}
\newcommand{\PLUSOp}{\mathsf{PLUS}}
\newcommand{\PLUS}[1]{\PLUSOp(#1)}
\newcommand{\TIMESOp}{\mathsf{TIMES}}
\newcommand{\TIMES}[1]{\TIMESOp(#1)}
\newcommand{\vlist}[1]{\mathit{#1}}
\newcommand{\fold}{\mathsf{fold}}
\newcommand{\CFGsubstOp}{\mathsf{close}}
\newcommand{\CFGsubst}{\mathsf{close}}
\newcommand{\In}{\keyw{in}}
\newcommand{\Nonterm}{\mathsf{NT}}
\newcommand{\Term}{\mathsf{T}}

Tries are not restricted to regular operations. We define an operation that
transforms a context-free grammar (in a particular normal form) into the
 trie denoting the same context-free language. The operation borrows ideas from Winter's et al.~\cite{WinterBR11} coalgebraic
exposition of context-free languages.

For the rest of the subsection, we fix a context-free grammar given by the type of its terminal symbols $\CFGTT$, the type of its non-terminal symbols $\CFGNT$, a distinguished starting non-terminal $\CFGS \COLON \CFGNT$, and its productions $\CFGP \COLON \CFGNT \to \setT{\listT{(\CFGTT + \CFGNT)}}$---an assignment of a set of words consisting of terminal and non-terminal symbols for each non-terminal. For readability, we write $\Term$ for $\Inl \COLON \CFGTT \to \CFGTT  + \CFGNT$ and $\Nonterm$ for $\Inr \COLON \CFGNT \to \CFGTT  + \CFGNT$ in this section. The inductive semantics of such a grammar is standard: a word over the alphabet $\CFGTT$ is accepted, if there is a finite derivation of that word using a sequence of productions. (For a formal definition, see Subsection~\ref{ssec:3:cfg}.)
For example, the grammar given by the productions
\begin{quotex}
$\kern-2em\CFGP\;N = \If\;N = \CFGS\;\Then\;\{\Nil,\,\List{\Term\;a},\,\List{\Term\;b},\,\List{\Term\;a,\,\Nonterm\;\CFGS,\,\Term\;a},\,
\List{\Term\;b,\,\Nonterm\;\CFGS,\,\Term\;b}\}\;\Else\;\{\}$
\end{quotex}
where $a,\, b \COLON \CFGTT$ with $a \not= b$, denotes the non-regular language of palindromes over $\{a,\,b\}$. In conventional syntax, we would write the above productions as $\mathsf{S} \to \epsilon \mid a \mid b \mid a\mathsf{S}a \mid b\mathsf{S}b$.

To give a coinductive semantics in form of a trie to a grammar, we must solve the word problem for context-free grammars and use that algorithm
to assign corecursively the Boolean labels in the trie. Different
algorithms solve the word problem for context-free grammars. Earley's algorithm~\cite{Earley70} is the most flexible (but also complex) one: it works for arbitrary grammars without requiring a syntactic normal form. To simplify the presentation, we work with a syntactic normal form that allows us to use a much simplified version of Earley's algorithm.
We require the productions to be in \emph{weak Greibach normal form}~\cite{WinterBR11}: every produced word should either be the empty list or start with a terminal. Formally:
\begin{quotex}
    $\forall N.\;\forall \alpha \in \CFGP\;N.\;\Case\;\alpha\;\Of\;\Nonterm\;M \ConsOp \_ \Rightarrow \False \mid \_ \Rightarrow \True$
\end{quotex}
{\looseness=-1
Weak Greibach normal form is a relaxation of the standard Greibach normal form~\cite{Greibach65}, which additionally requires the starting terminal to be followed only by non-terminals. The example palindrome grammar is in weak Greibach normal form but not in Greibach normal form.

}

{\setlooseness{-1}
The intermediate \emph{states} $\alpha,\,\beta,\,\ldots$ of a word derivation are words of type $\listT{(\CFGTT + \CFGNT)}$, which are reachable from the initial non-terminal. We observe that such a state can only produce the empty word according to $\CFGP$, if it consists only of non-terminals, each of which can immediately produce the empty word, i.e., $\Nil \in \CFGP\;N$. Note that due to weak Greibach normal form any non-$\Nil$ production will produce at least one terminal symbol.
We compute recursively whether a state can produce the empty word according to $\CFGP$ .
\begin{quotex}
    $\Primrec~\finalOp_\CFGP \COLON \listT{(\CFGTT + \CFGNT)} \to \boolT~\Where\\
    \TAB \finalOp_\CFGP\;\Nil = \True\\
    \TAB \finalOp_\CFGP\;(x \ConsOp \alpha) = \Case~x~\Of~\Term\;\_ \Rightarrow \False \mid \Nonterm\;N \Rightarrow \Nil \in \CFGP\;N \mathrel\land \finalOp_\CFGP\;\alpha$
\end{quotex}

}

A second useful function $\derivOp_\CFGP$ ``reads'' a terminal symbol $a$ in a state from the left, yielding a set of successor states to choose from non-deterministically. The recursive definition of $\derivOp_\CFGP$ is based on a similar observation as the one of $\finalOp_\CFGP$. If the state starts with a terminal $b$, the only successor state is the tail of the state if $a = b$. There is no successor state if $a \not = b$. If the state starts with a non-terminal $N$, we consider all non-empty productions in $\CFGP\;N$ starting with $a$ and replace $N$ with their tails. Additionally, if $N$ may produce the empty word, we drop it and continue recursively with the next terminal or non-terminal symbol of the state, which possibly results in additional successor states. Formally:
\begin{quotex}
    $\Primrec~\derivOp_\CFGP \COLON \listT{(\CFGTT + \CFGNT)} \to \CFGTT \to \setT{\listT{(\CFGTT + \CFGNT)}}~\Where\\
    \TAB \derivOp_\CFGP\;\Nil\;\_ = \{\}\\
    \TAB \derivOp_\CFGP\;(x \ConsOp \alpha)\;a = \Case~x~\Of~\\\TAB\TAB\phantom{{}\mid{}}\Term\;b \Rightarrow \If~a=b~\Then~\{\alpha\}~\Else~\{\}\\\TAB\TAB\mid \Nonterm\;N \Rightarrow
    \{\beta \AppendOp \alpha \mid (\Term\;a)\ConsOp\beta \in \CFGP\;N \} \mathrel\cup
    \If~\Nil \in \CFGP\;N~\Then~\derivOp_\CFGP\;\alpha\;a~\Else~\{\}$
\end{quotex}

Finally, we obtain a trie from a set of states by primitive corecursion using the two above functions to specify the observations. Note that the set of states changes when proceeding corecursively. For this definition we use the constructor view.
\begin{quotex}
    $\Primcorec~\CFGsubstOp \COLON \setT{\listT{(\CFGTT + \CFGNT)}} \to \langT{\CFGTT}~\Where\\
    \TAB \CFGsubst\;X = \LangC\;(\exists \alpha \in X.\;\finalOp_\CFGP\;\alpha)\;(\lambda a.\;\CFGsubst\;{\left(\bigcup_{\alpha \in X}\derivOp_\CFGP\;\alpha\;a\right))}$
\end{quotex}

The initial state, in which no terminal has been read yet, is the singleton list $[\CFGS]$. We obtain the trie $\mathsf{G}$ corresponding to our fixed grammar.
\begin{quotex}
    $\Definition~\mathsf{G} \COLON \langT{\CFGTT}~\Where\\
    \TAB \mathsf{G} = \CFGsubst\;\{[\CFGS]\}$
\end{quotex}

\subsection{Arbitrary Formal Languages}

Before we turn to proving, let us exercise one more corecursive definition.
Earlier, we have assigned each trie a set of lists via the function
$\fromLangOp$. Primitive corecursion enables us to define the converse operation.
\begin{quotex}
$\Primcorec~\toLangOp \COLON \setT{\listT{\alpha}} \to \langT{\alpha}~\Where\\
\TAB \final{(\toLang{L})} = \Nil \in L\\
\TAB \derivx{(\toLang{L})} = \lambda a.\;\toLang{\{w \mid \Cons{a}{w} \in L\}}$
\end{quotex}

The function $\fromLangOp$ and $\toLangOp$ are both bijections. We show this by
proving that their compositions (either way) are both the identity function. One
direction, $\fromLang{(\toLang{L})} = L$, follows by set extensionality and a
subsequent induction on words. The reverse direction requires a proof by
coinduction,  which is the topic of the next section.

Using $\toLangOp$ we can turn every (even undecidable) set of lists into a trie.
This is somewhat counterintuitive, since, given a concrete trie, its word
problem seems easily decidable (via the function $\inLangOp$). But of course in
order to compute the trie out of a set of lists $L$ via $\toLangOp$ the word
problem for $L$ must be solved---we are reminded that higher-order logic is not
a programming language where everything is executable, but a logic in which we
write down specifications (and not programs). %
For regular operations and context-free grammars from the previous
subsections the situation is different. For example, Isabelle's code generator
can produce valid Haskell code that evaluates
$\inLang{\Cons{a}{\Cons{b}{\Cons{a}{a}}}}{
\Star{(\Times{\Atomx{a}}{(\Plus{\Atomx{a}}{\Atomx{b}})})}}$ to $\True$ and $\inLang{\Cons{a}{\Cons{b}{\Cons{a}{a}}}}{\mathsf{G}}$ to $\False$, where $\mathsf{G}$ is the trie for the palindrome grammar from Subsection~\ref{subsec:cfg}. The latter is possible, despite the seemingly non-executable existential quantification and unions in the definition of $\CFGsubstOp$, due to Isabelle's code generator, which makes (co)finite sets executable through data refinement to lists or red-black trees~\cite{HaftmannKKN-ITP13}.

\section{Reasoning about Tries}
\label{sec:3:coind}

We have seen the definitions of many operations, justifying their meaningfulness
by appealing to the reader's intuition. This is often not enough to guarantee
correctness, especially if we have a theorem prover at hand. So let us formally
prove in Isabelle that the regular operations on tries form a Kleene algebra by
proving Kozen's famous axioms~\cite{Kozen94} as equalities (Subsection~\ref{ssec:3:coind}) or inequalities (Subsection~\ref{ssec:3:sim}) on tries and prove the soundness our our trie construction for a context free grammar with respect to the standard inductive semantics of grammars (Subsection~\ref{ssec:3:cfg}).

\subsection{Proving Equalities on Tries}
\label{ssec:3:coind}

Codatatypes are equipped with the perfect tool for proving equalities: the
coinduction principle. Intuitively, this principle states that the existence of
a relation $R$ that is \emph{closed} under the codatatype's observations (given
by selectors) implies that elements related by $R$ are equal. Being closed means
here that for all $R$-related codatatype elements their immediate
(non-corecursive) observations are equal and the corecursive observations are
again related by $R$. In other words, if we cannot distinguish elements of a
codatatype by (finite) observations, they must be equal. Formally, for our
codatatype $\langT{\alpha}$ we obtain the following coinduction rule. %
\vskip\abovedisplayskip
\begin{center}
\AXC{$R\;L\;K$}
\AXC{\quad$\forall L_1\;L_2.\;R\;L_1\;L_2 \RIMP (\final{L_1} \IFF \final{L_2} \LAND \forall x.\;
  R\;(\deriv{L_1}{x})\;(\deriv{L_2}{x}))$}
\RightLabel{$\langCI$}
\BIC{$L = K$}
\DP
\end{center}
\vskip\belowdisplayskip
The second assumption of this rule formalizes the notion of being closed under
observations: If two tries are related then their root's labels are
the same and all their subtrees are related. A relation that satisfies this
assumption is called a \emph{bisimulation}. Thus, proving an equation by
coinduction amounts to exhibiting a bisimulation witness that relates the left
and the right hand sides.

Let us observe the coinduction rule, which we call $\langCI$, in action. Figure~\ref{fig-proof} shows a detailed proof of the Kleene algebra axiom
that the empty language is the left identity of union that is accepted by Isabelle.%
\begin{figure}
\begin{quotex}
{\small1}\TAB$\Theorem~\Plus{\Zero}{L} = L$\\
\noindent{\small2}\TAB$\IsarProof~(\Rule~\langCI)$\\
\noindent{\small3}\TAB$\TAB \Def~R~L_1~L_2 = (\exists K.~L_1 = \Plus{\Zero}{K} \LAND L_2 = K)$\\[2\jot]
\noindent{\small4}\TAB$\TAB \Show~R~(\Plus{\Zero}{L})~L~\By~\Simp$\\[2\jot]
\noindent{\small5}\TAB$\TAB \Fix~L_1~L_2$\\
\noindent{\small6}\TAB$\TAB \Assume~R~L_1~L_2$\\
\noindent{\small7}\TAB$\TAB \Then~\Obtain~K~\Where~L_1 = \Plus{\Zero}{K}~\AND~L_2 = K~\By~\Simp$\\
\noindent{\small8}\TAB$\TAB \Then~\Show~\final{L_1} \IFF \final{L_2} \LAND \forall x.\;
  R\;(\deriv{L_1}{x})\;(\deriv{L_2}{x})~\By~\Simp$\\
\noindent{\small9}\TAB$\kQed$
\end{quotex}
\caption{A detailed proof by coinduction}
\label{fig-proof}
\end{figure}
After applying the coinduction rule backwards (line~2), the proof has three
parts. First, we supply a definition of a witness relation $R$ (line~3). Second, we prove
that $R$ relates the left and the right hand side (line~4). Third, we prove that $R$ is a
bisimulation (lines~5--8). Proving $R~(\Plus{\Zero}{L})~L$ for
our particular definition of $R$ is trivial after instantiating the
existentially quantified $K$ with $L$. Proving the bisimulation property is
barely harder: for two tries $L_1$ and $L_2$ related by $R$ we need to show
$\final{L_1} \IFF \final{L_2}$ and $\forall x.\;
R\;(\deriv{L_1}{x})\;(\deriv{L_2}{x})$. Both properties follow by simple
calculations rewriting $L_1$ and $L_2$ in terms of a trie $K$ (line 7), whose existence
is guaranteed by $R~L_1~L_2$, and simplifying with the selector equations for
$\PlusOp$ and $\Zero$. %
\begin{quotex}
$\final{L_1} \IFF \final{(\Plus{\Zero}{K})} \IFF (\final{\Zero} \LOR \final{K}) \IFF
(\False \LOR \final{K}) \IFF \final{K} \IFF \final{L_2}$\\[2\jot]
$R\;(\deriv{L_1}{x})\;(\deriv{L_2}{x}) \IFF
 R\;(\deriv{(\Plus{\Zero}{K})}{x})\;(\deriv{K}{x}) \\\IFF
 R\;(\Plus{\deriv{\Zero}{x}}{\deriv{K}{x}})\;(\deriv{K}{x}) \IFF
 R\;(\Plus{\Zero}{\deriv{K}{x}})\;(\deriv{K}{x}) \\\IFF
(\exists K'.~\Plus{\Zero}{\deriv{K}{x}} =\Plus{\Zero}{K'} \LAND \deriv{K}{x} = K') \IFF \True$
\end{quotex}
The last step is justified by instantiating $K'$ with $\deriv{K}{x}$.

So in the end, the only part that required ingenuity was the definition of the witness $R$. But was it truly ingenious?
It turns out that in general, when proving a conditional equation
$P\;\overline{x} \RIMP l\;\overline{x} = r\;\overline{x}$ by coinduction, where
$\overline{x}$ denotes a list of variables occurring freely in the equation, the
canonical choice for the bisimulation witness is $R\;a\;b =\allowbreak
(P\;\overline{x} \LAND \exists\overline{x}.~a = l\;\overline{x} \LAND b = r\;\overline{x})$. There is no guarantee that this definition yields a
bisimulation. However, after performing a few proofs by coinduction, this
particular pattern emerges as a natural first choice to try.
Indeed, the choice is so natural, that it was worth to automate it in the
$\Coinduction$ proof method~\cite{BlanchetteHLPPT-ITP14}. This method
instantiates the coinduction rule $\langCI$ with the canonical bisimulation
witness constructed from the goal, where the existentially quantified variables
must be given explicitly using the $\Arbitrary$ modifier. 
Then it applies the instantiated rule in a resolution step, discharges the first
assumption, and unpacks the existential quantifiers from $R$ in the remaining
subgoal (the $\Obtain$ step in the above proof).
Many proofs collapse to an automatic one-line proof using this convenience, including the one above. Some
examples follow. %
\begin{quotex}
$\Theorem~\Plus{\Zero}{L} = L~~$
$\By~(\Coinduction~\ArbitraryMod~L)~\Auto$\\[2\jot]
$\Theorem~\Plus{L}{L} = L~~$
$\By~(\Coinduction~\ArbitraryMod~L)~\Auto$\\[2\jot]
$\Theorem~\Plus{L_1}{L_2} = \Plus{L_2}{L_1}~~$
$\By~(\Coinduction~\ArbitraryMod~L_1~L_2)~\Auto$\\[2\jot]
$\Theorem~\Plus{(\Plus{L_1}{L_2})}{L_3} = \Plus{L_1}{\Plus{L_2}{L_3}}~~$
$\By~(\Coinduction~\ArbitraryMod~L_1~L_2~L_3)~\Auto$\\[2\jot]
$\Theorem~\toLang{(\fromLang{L})} = L~~$
$\By~(\Coinduction~\ArbitraryMod~L)~\Auto$\\[2\jot]
$\Theorem~\toLang{(L \CUP K)} = \Plus{\toLang{L}}{\toLang{K}}~~$
$\By~(\Coinduction~\ArbitraryMod~L~K)~\Auto$
\end{quotex}

As indicated earlier, sometimes the $\Coinduction$ method does not succeed. It
is instructive to consider one example where this is the case:
$\final{L} \RIMP \Plus{\One}{L} = L$.

If we attempt to prove the above statement by coinduction instantiated with the
canonical bisimulation witness $R~L_1~L_2 = (\exists K.~L_1 = \Plus{\One}{K}
\LAND L_2 = K \LAND \final{K})$, after some simplification we are stuck with
proving that $R$ is a bisimulation. %
\begin{quotex}
$R\;(\deriv{L_1}{x})\;(\deriv{L_2}{x}) \IFF
 R\;(\deriv{(\Plus{\One}{K})}{x})\;(\deriv{K}{x}) \\\IFF
 R\;(\Plus{\deriv{\One}{x}}{\deriv{K}{x}})\;(\deriv{K}{x}) \IFF
 R\;(\Plus{\Zero}{\deriv{K}{x}})\;(\deriv{K}{x}) \\\IFF
 R\;(\deriv{K}{x})\;(\deriv{K}{x}) \IFF (\exists K'.~\deriv{K}{x} =\Plus{\One}{K'} \LAND \deriv{K}{x} = K' \LAND \final{K'})$
\end{quotex}

The remaining goal is not provable: we would need to instantiate $K'$ with
$\deriv{K}{x}$, but then, we are left to prove $\final{(\deriv{K}{x})}$, which
we cannot deduce (we only know $\final{K}$). If we, however, instantiate the
coinduction rule with $\reflCl{R}~L_1~L_2 = R~L_1~L_2 \LOR L_1 = L_2$, we are
able to finish the proof. This means that our canonical bisimulation witness $R$
is not a bisimulation, but $\reflCl{R}$ is. In such cases $R$ is called a
\emph{bisimulation up to equality}~\cite{Sangiorgi98}.

Instead of modifying the $\Coinduction$ method to instantiate the rule $\langCI$ with
$\reflCl{R}$, it is more convenient to capture this improvement directly in the
coinduction rule. This results in the following rule which we call
\emph{coinduction up to equality} or $\langCIeq$. %
\vskip\abovedisplayskip
\begin{center}
\AXC{$R\;L\;K$}
\AXC{\quad$\forall L_1\;L_2.\;R\;L_1\;L_2 \RIMP (\final{L_1} \IFF \final{L_2} \LAND \forall x.\;
  \reflCl{R}\;(\deriv{L_1}{x})\;(\deriv{L_2}{x}))$}
\RightLabel{$\langCIeq$}
\BIC{$L = K$}
\DP
\end{center}
\vskip\belowdisplayskip

Note that $\langCIeq$ is not just an instance of $\langCI$, with $R$ replaced by
$\reflCl{R}$. Instead, after performing this replacement, the first assumption
is further simplified to $R~L~K$---we would not use coinduction in the first
place, if we could prove $\reflCl{R}~L~K$ by reflexivity. Similarly, the
reflexivity part in the occurrence on the left of the implication in the second
assumption is needless and therefore eliminated. The resulting coinduction up to
equality principles are independent of the particular codatatypes and thus
uniformly produced by the \keyw{codatatype} command. Using this coinduction up
to equality rule, we have regained the ability to write one-line proofs. %
\begin{quotex}
$\Theorem~\final{L} \RIMP \Plus{\One}{L} = L~~$
$\By~(\Coinduction~\ArbitraryMod~L~\RuleMod~\langCIeq)~\Auto$
\end{quotex}

One might think that the principle $\langCIeq$ is always preferable to $\langCI$. This is true from the expressiveness point of view: whatever can be proved with $\langCI$, can also be proved with $\langCIeq$. However, for proof automation $\langCIeq$ is often less beneficial: to prove membership in $\reflCl{R}$ we need to prove a disjunction which may result in a larger search space, given that neither of the disjuncts is trivially false. In summary, using the weakest rule that suffices to finish the proof helps proof automation.

This brings us to the next hurdle. Suppose that we already have
proved the natural selector equations~\eqconc{} for concatenation $\TimesOp$. (This
requires finding some auxiliary properties of $\TimesLROp$ and $\TimesPlusOp$
but is an overall straightforward usage of coinduction up to equality.) Next,
we want to reason about $\TimesOp$. For example, we prove its distributivity
over $\PlusOp$: $\Times{(\Plus{L}{K})}{M} =
\Plus{(\Times{L}{M})}{(\Times{K}{M})}$. As before, we are stuck proving that the
canonical bisimulation witness $R~L_1~L_2 = (\exists L'~K'~M'.~L_1 =
\Times{(\Plus{L'}{K'})}{M'} \LAND L_2 =
\Plus{(\Times{L'}{M'})}{(\Times{K'}{M'})})$ is a bisimulation (and this time
even for up to equality). %
\begin{quotex}
$\reflCl{R}\;(\deriv{L_1}{x})\;(\deriv{L_2}{x}) \IFF
 \reflCl{R}\;(\deriv{(\Times{(\Plus{L'}{K'})}{M'})}{x})\;(\deriv{(\Plus{(\Times{L'}{M'})}{(\Times{K'}{M'})})}{x}) \kern-2em\\[1\jot]\IFF
  \begin{cases}
    \reflCl{R}\;(\Times{(\Plus{\deriv{L'}{x}}{\deriv{K'}{x}})}{M'})\\
    \phantom{\reflCl{R}}\;(\Plus{(\Times{\deriv{L'}{x}}{M'})}{(\Times{\deriv{K'}{x}}{M'})})&\If~ \NOT \final{L'} \LAND \NOT \final{K'}\\
   \reflCl{R}\;(\Plus{\Times{(\Plus{\deriv{L'}{x}}{\deriv{K'}{x}})}{M'}}{\deriv{M'}{x}})\\
     \phantom{\reflCl{R}}\;(\Plus{(\Plus{\Times{\deriv{L'}{x}}{M'}}{\deriv{M'}{x}})}{(\Times{\deriv{K'}{x}}{M'})})&\If~\final{L'} \LAND \NOT \final{K'}\\
   \reflCl{R}\;(\Plus{\Times{(\Plus{\deriv{L'}{x}}{\deriv{K'}{x}})}{M'}}{\deriv{M'}{x}})\\
   \phantom{\reflCl{R}}\;(\Plus{(\Times{\deriv{L'}{x}}{M'})}{(\Plus{\Times{\deriv{K'}{x}}{M'}}{\deriv{M'}{x}})})&\If~\NOT \final{L'} \LAND \final{K'}\\
   \reflCl{R}\;(\Plus{\Times{(\Plus{\deriv{L'}{x}}{\deriv{K'}{x}})}{M'}}{\deriv{M'}{x}})\\
   \phantom{\reflCl{R}}\;(\Plus{(\Plus{\Times{\deriv{L'}{x}}{M'}}{\deriv{M'}{x}})}{(\Plus{\Times{\deriv{K'}{x}}{M'}}{\deriv{M'}{x}})})&\If~\final{L'} \LAND \final{K'}
  \end{cases}\kern-3em\\[1\jot]\IFF
   \begin{cases}
     \True&\If~ \NOT \final{L'} \LAND \NOT \final{K'}\\
    \reflCl{R}\;(\Plusg{\Times{(\Plus{\deriv{L'}{x}}{\deriv{K'}{x}})}{M'}}{\deriv{M'}{x}})\\
    \phantom{\reflCl{R}}\;(\Plusg{(\Plus{\Times{\deriv{L'}{x}}{M'}}{\Times{\deriv{K'}{x}}{M'}})}{\deriv{M'}{x}})&\text{otherwise}
   \end{cases}\kern-2em\\$
\end{quotex}

The remaining subgoal cannot be discharged by the definition of $R$. In
principle it could be discharged by equality (the two tries are equal), but this
is almost exactly the property we originally started proving, so we have not
simplified the problem by coinduction but rather are going in circles here.
However, if our relation could somehow split its arguments across the outermost
$\PlusOp$ highlighted in gray, we could prove the left pair being related by $R$ and the right pair
by $=$. The solution is easy: we allow the relation to do just that. Accordingly,
we define the congruence closure $\PlusCl{R}$ of a relation $R$ under $\PlusOp$
inductively by the following rules.  %

\begin{quotex}\leavevmode\kern-3em
\begin{tabular}{@{}c@{\qquad}c@{\qquad}c@{\qquad}c@{\qquad}c@{}}
\AXC{$L = K\vphantom{\PlusCl{R}}$}
\UIC{$\PlusCl{R}\;L\;K$}
\DP
&
\AXC{$R\;L\;K\vphantom{\PlusCl{R}}$}
\UIC{$\PlusCl{R}\;L\;K$}
\DP
&
\AXC{$\PlusCl{R}\;L\;K$}
\UIC{$\PlusCl{R}\;K\;L$}
\DP
&
\AXC{$\PlusCl{R}\;L_1\;L_2$}
\AXC{$\PlusCl{R}\;L_2\;L_3$}
\BIC{$\PlusCl{R}\;L_1\;L_3$}
\DP
&
\AXC{$\PlusCl{R}\;L_1\;K_1$}
\AXC{$\PlusCl{R}\;L_2\;K_2$}
\BIC{$\PlusCl{R}\;(\Plus{L_1}{L_2})\;(\Plus{K_1}{K_2})$}
\DP
\end{tabular}\kern-1em
\end{quotex}

The closure $\PlusCl{R}$ is then used to strengthen the coinduction rule, just
as the earlier reflexive closure $\reflCl{R}$ strengthening. Note that the
closure $\reflCl{R}$ can also be viewed as inductively generated by the first
two of the above rules. In summary, we obtain \emph{coinduction up to congruence of
$\PlusOp$}, denoted by $\langCIplus$. %
\vskip\abovedisplayskip
\begin{center}
\AXC{$R\;L\;K$}
\AXC{$\forall L_1\;L_2.\;R\;L_1\;L_2 \RIMP (\final{L_1} \IFF \final{L_2} \LAND \forall x.\;
  \PlusCl{R}\;(\deriv{L_1}{x})\;(\deriv{L_2}{x}))$}
\RightLabel{$\langCIplus$}
\BIC{$L = K$}
\DP
\end{center}
\vskip\belowdisplayskip
This rule is easily derived from plain coinduction by instantiating $R$ in $\langCI$ with $\PlusCl{R}$ and proceeding by induction on the definition of the congruence closure.

{\setlooseness{-1}
As intended $\langCIplus$ makes the proof of distributivity into another
automatic one-liner. This is because our new proof principle,
coinduction up to congruence of $\PlusOp$, matches exactly the definitional
principle of corecursion up to $\PlusOp$ used in the selector equations~\eqconc{} of~$\TimesOp$. %
\begin{quotex}
$\Theorem~\Times{(\Plus{L}{K})}{M} = \Plus{(\Times{L}{M})}{(\Times{K}{M})}$\\
$\TAB \By~(\Coinduction~\ArbitraryMod~L~K~M~\RuleMod~\langCIplus)~\Auto$
\end{quotex}

}

Coinduction up to congruence of $+$ allows us also to prove properties of the shuffle product, e.g., commutativity $\Shuffle{L}{K} = \Shuffle{K}{L}$ and associativity $\Shuffle{(\Shuffle{K}{L})}{M} = \Shuffle{K}{(\Shuffle{L}{M})}$.

For properties involving iteration $\StarOp$, whose selector equations~\eqstar{} are
corecursive up to~$\TimesOp$, we will need a further strengthening of the
coinduction rule (using the congruence closure under $\TimesOp$). Overall, the
most robust solution to keep track of the different rules is to maintain a
coinduction rule up to all previously defined operations on tries: we define
$\regCl{R}$ to be the congruence closure of $R$ under $\PlusOp$, $\TimesOp$, and
$\StarOp$ and then use the following rule for proving. %
\vskip\abovedisplayskip
\begin{center}
\AXC{$R\;L\;K$}
\AXC{$\forall L_1\;L_2.\;R\;L_1\;L_2 \RIMP (\final{L_1} \IFF \final{L_2} \LAND \forall x.\;
  \regCl{R}\;(\deriv{L_1}{x})\;(\deriv{L_2}{x}))$}
\RightLabel{$\langCIreg$}
\BIC{$L = K$}
\DP
\end{center}
\vskip\belowdisplayskip

We will not spell out all equational axioms of Kleene algebra~\cite{Kozen94} and their
formal proofs~\cite{Traytel-AFP13} here. Most proofs are automatic; some require
a few manual hints in which order to apply the congruence rules.

\subsection{Proving Inequalities on Tries}
\label{ssec:3:sim}

A few axioms of Kleene algebra also contain inequalities, such as
$\Plus{\One}{\Times{L}{\Star{L}}} \leq \Star{L}$, and even conditional
inequalities, such as $\Plus{L}{\Times{M}{K}} \leq M \RIMP \Times{L}{\Star{K}}
\leq M$. On languages, $L \leq K$ is defined as $\Plus{L}{K} = K$, such that in principle proofs by
coinduction also are applicable here. However, we can achieve even better proof automation, if we formulate and use the following dedicated coinduction principle for $\leq$.
\vskip\abovedisplayskip
\begin{center}
\AXC{$R\;L\;K$}
\AXC{$\forall L_1\;L_2.\;R\;L_1\;L_2 \RIMP ((\final{L_1} \RIMP \final{L_2}) \LAND \forall x.\;
R\;(\deriv{L_1}{x})\;(\deriv{L_2}{x}))$}
\RightLabel{$\lessCI$}
\BIC{$L \leq K$}
\DP
\end{center}
\vskip\belowdisplayskip
This theorem has the same shape as the usual coinduction principle $\langCI$, however the relation $R$ is only required to be a simulation instead of a bisimulation. In other words $R$ still needs to be closed under corecursive observations, however the immediate observation of the first argument must only imply the one of the second argument (as opposed to being equal to it). We call this coinduction principle $\lessCI$ and prove it by unfolding the definition of $\leq$ and then using the equational coinduction principle $\langCI$.

While $\lessCI$ allows us to prove simple properties like $r \leq \Plus{r}{s}$, it is not strong enough to automatically prove the inequational Kleene algebra axioms, which involve concatenation and iteration. As in the case of equations, up-to reasoning is the familiar way out of this dilemma. However, since $R$ is only a simulation, and thus in general not an equivalence relation, we can not consider its congruence closure. Instead, we follow Rot et al.~\cite{RotBR16} and define inductively the so-called precongruence closure $\PlusClsim{R}$\,.

\begin{quotex}
    \begin{tabular}{@{}c@{\qquad}c@{\qquad}c@{\qquad}c@{}}
        \AXC{$L \leq K\vphantom{\PlusClsim{R}}$}
        \UIC{$\PlusClsim{R}\;L\;K$}
        \DP
        &
        \AXC{$R\;L\;K\vphantom{\PlusClsim{R}}$}
        \UIC{$\PlusClsim{R}\;L\;K$}
        \DP
        &
        \AXC{$\PlusClsim{R}\;L_1\;L_2$}
        \AXC{$\PlusClsim{R}\;L_2\;L_3$}
        \BIC{$\PlusClsim{R}\;L_1\;L_3$}
        \DP
        &
        \AXC{$\PlusClsim{R}\;L_1\;K_1$}
        \AXC{$\PlusClsim{R}\;L_2\;K_2$}
        \BIC{$\PlusClsim{R}\;(\Plus{L_1}{L_2})\;(\Plus{K_1}{K_2})$}
        \DP
        \\\\
        \multicolumn{2}{@{}c@{\quad}}{
        \AXC{$\PlusClsim{R}\;L_1\;K_1$}
        \AXC{$\PlusClsim{R}\;L_2\;K_2$}
        \BIC{$\PlusClsim{R}\;(\Times{L_1}{L_2})\;(\Times{K_1}{K_2})$}
        \DP}
        &
        \AXC{$\PlusClsim{R}\;L\;K$}
        \UIC{$\PlusClsim{R}\;(\Star{L})\;(\Star{K})$}
        \DP
        &
        \AXC{$\PlusClsim{R}\;L_1\;K_1$}
        \AXC{$\PlusClsim{R}\;L_2\;K_2$}
        \BIC{$\PlusClsim{R}\;(\Inter{L_1}{L_2})\;(\Inter{K_1}{K_2})$}
        \DP
    \end{tabular}\kern-1em
\end{quotex}

With this definition we are able to prove the following strengthened coinduction principle up to precongruence closure, called $\lessCIreg$.
\vskip\abovedisplayskip
\begin{center}
\AXC{$R\;L\;K$}
\AXC{$\forall L_1\;L_2.\;R\;L_1\;L_2 \RIMP ((\final{L_1} \RIMP \final{L_2}) \LAND \forall x.\;
    \PlusClsim{R}\;(\deriv{L_1}{x})\;(\deriv{L_2}{x}))$}
\RightLabel{$\lessCIreg$}
\BIC{$L \leq K$}
\DP
\end{center}
\vskip\belowdisplayskip
The proof of $\lessCIreg$ is structurally very similar to the one of $\langCIreg$: after using the plain $\lessCI$ rule, we are left with proving that the precongruence closure $\PlusClsim{R}$ is a simulation. This follows by induction on the definition of the preconguence closure.
Crucially, the complement operation $\NotOp$ is not included in this definition. For simulation up-to preconguence closure to be a simulation, all operations must be monotone with respect to their immediate observations $\finalOp$, which is not the case for $\NotOp$~\cite{RotBR16}.

Finally, we are capable to write automatic proofs for inequalities.

\begin{quotex}
    $\Theorem~\Plus{\One}{\Times{L}{\Star{L}}} \leq \Star{L}~~$
    $\By~(\Coinduction~\RuleMod~\lessCIreg)~\Auto$\\[2\jot]
    $\Theorem~\final{K} \RIMP L \leq \Times{L}{K}~~$
    $\By~(\Coinduction~\ArbitraryMod~L~\RuleMod~\lessCIreg)~\Auto$
\end{quotex}

We remark that working with inequalities also has its cost. The reason for this is that Isabelle excels at equational reasoning. Isabelle also provides automation for reasoning with orders, but it is noticeably less powerful than the one for $=$. On the other hand, equations of the form $\Plus{L}{K} = K$, which one gets after unfolding the definition of $\leq$, are not ideal for rewriting. Proofs that reduce inequalities to equalities ofter require manual hints to expand $K$ into $\Plus{L}{K}$ at the right places. In the end, when using up-to simulations a careful setup of rewriting rules and classical reasoning support for $\leq$ results in a higher degree of automation. This is especially perceivable for more complicated inequational properties like $\Plus{L}{\Times{M}{K}} \leq M \RIMP \Times{L}{\Star{K}}
\leq M$.

\subsection{Reasoning about Context-Free Languages}
\label{ssec:3:cfg}

We connect our trie construction from Subsection~\ref{subsec:cfg} for a context-free grammar in weak Greibach normal form to the traditional inductive semantics of context-free grammars. We use the notational conventions and definition of Subsection~\ref{subsec:cfg}, including fixing the starting non-terminal $\CFGS \COLON \CFGNT$ and the productions $\CFGP \COLON \CFGNT \to \setT{\listT{(\CFGTT + \CFGNT)}}$. First, we formalize the traditional inductive semantics using an inductive binary predicate $\inLangOp_\CFGP \COLON \listT{\CFGTT} \to \listT{(\CFGTT + \CFGNT)} \to \boolT$ (written infix). Intuitively, $w \mathrel{\inLangOp_\CFGP} \alpha$ holds if and only if $w$ is derivable from $\alpha$ in finitely many production steps via $\mathsf{P}$, where each time we replace the leftmost non-terminal first.
\vskip\abovedisplayskip
\begin{center}
\begin{tabular}{@{}c@{\qquad}c@{\qquad}c@{}}
\AXC{$\Nil \mathrel{\inLangOp_\CFGP} \Nil$}
\DP&
\AXC{$w \mathrel{\inLangOp_\CFGP} \alpha$}
\UIC{$(a \ConsOp w) \mathrel{\inLangOp_\CFGP} (\Term\;a \ConsOp \alpha)$}
\DP&
\AXC{$\exists \beta \in \CFGP\;N.\;w \mathrel{\inLangOp_\CFGP} \beta\AppendOp\alpha$}
\UIC{$w \mathrel{\inLangOp_\CFGP} (\Nonterm\;N \ConsOp \alpha)$}
\DP
\end{tabular}
\end{center}
\vskip\belowdisplayskip
Note that $\inLangOp_\CFGP$ gives a way to assign the language $\{w \mid w \mathrel{\inLangOp_\CFGP} [N]\}$ to each non-terminal $N$, and in particular the language $\{w \mid w \mathrel{\inLangOp_\CFGP} [\CFGS]\}$ for the whole grammar given by $\CFGP$ and $\CFGS$.
We now prove that our trie $\CFG$ for the fixed grammar represents the same language, i.e., $\CFG = \toLang{\{w \mid w \mathrel{\inLangOp_\CFGP} [\CFGS]\}}$. Our proof uses an auxiliary intermediate inductive predicate $\inLangOp^{\finalOp\derivOp}_\CFGP \COLON \listT{\CFGTT} \to \setT{\listT{(\CFGTT + \CFGNT)}} \to \boolT$ (written infix) that reflects the change of the set of states during corecursion in $\CFGsubst$ function (which is used to construct $\CFG$).
\vskip\abovedisplayskip
\begin{center}
\begin{tabular}{@{}c@{\qquad}c@{\qquad}c@{}}
\AXC{$\exists \alpha \in X.\;\finalOp_\CFGP\;\alpha$}
\UIC{$\Nil \mathrel{\inLangOp^{\finalOp\derivOp}_\CFGP} X$}
\DP&
\AXC{$w \mathrel{\inLangOp^{\finalOp\derivOp}_\CFGP} (\bigcup_{\alpha \in X}.\;\derivOp_\CFGP\;\alpha\;a)$}
\UIC{$(a \ConsOp w) \mathrel{\inLangOp^{\finalOp\derivOp}_\CFGP} X$}
\DP
\end{tabular}
\end{center}
\vskip\belowdisplayskip
In some sense, $\inLangOp^{\finalOp\derivOp}_\CFGP$ is the inductive view on the $\CFGsubstOp$ function, as established next.
\begin{quotex}
$\Theorem~\CFGsubst\;X = \toLang{\{w\mid w \mathrel{\inLangOp^{\finalOp\derivOp}_\CFGP} X\}}
~\By~(\Coinduction~\ArbitraryMod~X)~\Auto$
\end{quotex}
The proof uses the simplest coinduction rule $\langCI$ and relies on injectivity of $\toLangOp$.

Next, we establish that $w\mathrel{\inLangOp_\CFGP} \alpha$ holds if and only if $w\mathrel{\inLangOp^{\finalOp\derivOp}_\CFGP} \{\alpha\}$ holds. We prove the two directions separately. Thereby we generalize the ``if''-direction.
\begin{quotex}
$\Theorem~w \mathrel{\inLangOp^{\finalOp\derivOp}_\CFGP} X \RIMP \exists \alpha \in X.\;w\mathrel{\inLangOp_\CFGP} \alpha$\\
$\Theorem~w\mathrel{\inLangOp_\CFGP} \alpha \RIMP w \mathrel{\inLangOp^{\finalOp\derivOp}_\CFGP} \{\alpha\}$
\end{quotex}
We do not show the proofs for the above statements about since both are standard inductions on the inductive definitions of $\inLangOp^{\finalOp\derivOp}_\CFGP$ and $\inLangOp_\CFGP$. Out of all theorems shown in this subsection, only the last one requires the grammar to be in weak Greibach normal form. Putting the three theorems together, we obtain the desired characterization: $\CFG = \toLang{\{w \mid w \mathrel{\inLangOp_\CFGP} [\CFGS]\}}$.

We remark that a definition of the trie $\CFG$ via a full Earley parser, would not only remove the need of weak Greibach normal form, but also enable coinductive reasoning about arbitrary grammars. For example, one could also formalize and prove correct the translation of grammars to certain normal forms without resorting to the traditional inductive semantics (and thus induction). We leave this study as future work, as it may burst the scope of an introduction to coinduction.

\section{Coalgebraic Foundations}
\label{sec:3:coalg}

We briefly connect the formalized but still intuitive notions, such as tries,
from earlier sections with the key coalgebraic concepts and terminology that is
usually used to present the coalgebraic view on formal languages. Thereby, we
explain how particularly useful abstract objects gave rise to concrete tools in
Isabelle\slash HOL. More theoretical and detailed introductions to coalgebra can
be found elsewhere~\cite{Rutten00,Jacobs1997}.

Given a functor $\ty{F}$ an ($\ty{F}$-)\emph{coalgebra} is a \emph{carrier} object
$A$ together with a map $A \to \ty{F}\;A$---the \emph{structural map} of a
coalgebra. In the context of higher-order logic---that is in the category of
types which consists of types as objects and of functions
between types as arrows---a functor is a type constructor $\ty{F}$ together
with a map function $\map_\ty{F} \COLON (\alpha \to \beta) \to \alpha\;\ty{F}
\to \beta\;\ty{F}$ that preserves identity and composition: $\map_\ty{F}\;\Id =
\Id$ and $\map_\ty{F}\;(f \circ g) = \map_\ty{F}\;f\circ \map_\ty{F}\;g$. An
$\ty{F}$-coalgebra in HOL is therefore simply a function $s \COLON \alpha \to
\alpha\;\ty{F}$. A function $f \COLON \alpha \to \beta$ is a coalgebra
\emph{morphism} between two coalgebras $s \COLON \alpha \to \alpha\;\ty{F}$ and
$t \COLON \beta \to \beta\;\ty{F}$ if it satisfies the commutation property $t \circ
f = \map_\ty{F}\;f \circ s$, also depicted by the commutative diagram in
Figure~\ref{fig:3:morphism}.

\begin{figure}[t]
\centering
$
\xymatrix@C=3pc@R=1pc{
\alpha
\ar^{\hspace*{-1ex}s}[r]
\ar_{f}[d]    &
\alpha\;\ty{F}
\ar^{\map_\ty{F}\;f}[d]  \\
\beta
\ar^{\hspace*{-1ex}t}[r]  & \beta\;\ty{F}
}
$
\caption{Commutation property of a coalgebra morphism}
\label{fig:3:morphism}
\end{figure}

{
\looseness=-1
An ($\ty{F}$-)coalgebra to which there exists a unique morphism from any other
coalgebra is called a \emph{final} ($\ty{F}$-)\emph{coalgebra}. Not all functors
$\ty{F}$ admit a final coalgebra~\cite[Section~10]{Rutten00}. Two different final coalgebras are necessarily
isomorphic. Final coalgebras correspond to codatatypes in Isabelle\slash HOL.
Isabelle's facility for defining codatatypes maintains a large class of
functors---bounded natural functors~\cite{TraytelPB-LICS12}---for which a final
coalgebra does exists. Moreover, for any bounded natural functor $\ty{F}$,
Isabelle can construct its final coalgebra with the codatatype $\ty{CF}$ as the
carrier and define a bijective constructor $\const{C_\ty{F}} \COLON
\ty{CF}\;\ty{F} \to \ty{CF}$ and its inverse, the destructor $\const{D_\ty{F}} \COLON
\ty{CF} \to \ty{CF}\;\ty{F}$. The latter takes the role of the structural map of
the coalgebra. %
\begin{quotex}
$\Codatatype~\ty{CF} = \const{C_\ty{F}}~(\const{D_\ty{F}}:\ty{CF}\;\ty{F})$
\end{quotex}

}

Finally, we are ready to connect these abstract notions to our tries. The
codatatype of tries $\langT{\alpha}$ is the final coalgebra of the functor
$\beta\;\ty{D} = \boolT \times (\alpha \to \beta)$ with the associated map
function $\map_\ty{D}\;g = \Id \times (\lambda f.\;g\circ f)$, where $(f \times
g)\;(x,\,y) = (f\;x,\,g\;y)$. The structural map of this final coalgebra is the
function $\const{D_\ty{D}} = \langle\finalOp,\,\derivOp\rangle$, where $\langle
f,\; g\rangle\;x = (f\;x,\,g\;x)$.

The finality of $\langT{\alpha}$ gives rise to the definitional principles of
primitive coiteration and corecursion. In Isabelle the coiteration principle is embodied by the primitive
\emph{coiterator} $\coiter \COLON (\tau \to \tau\;\ty{D}) \to \tau \to \langT{\alpha}$,
that assigns to the given $\ty{D}$-coalgebra the unique morphism from itself to the
final coalgebra. In other words, the primitive coiterator allows us to define
functions of type $\tau \to \langT{\alpha}$ by providing a $\ty{D}$-coalgebra
on $\tau$, i.e., a function of type $\tau \to \boolT \times (\alpha \to \tau)$
that essentially describes a deterministic (not necessarily finite) automaton
without an initial state. To clarify this automaton analogy, it is customary to
present the $\ty{F}$-coalgebra $s$ as two functions $s = \langle \const{o},\,
\const{d} \rangle$ with $\tau$ being the states of the automaton, $\const{o} :
\tau \to \boolT$ denoting accepting states, and $\const{d} : \alpha \to \tau \to
\tau$ being the transition function. From a given $s$, we uniquely obtain the
function $\coiter\;s$ that assigns to a separately given initial state $t : \tau$
the language $\coiter\;s\;t : \langT{\alpha}$ and makes the diagram in
Figure~\ref{fig-coiter} commute. Note that Figure~\ref{fig-coiter} is an instance of Figure~\ref{fig:3:morphism}.

\begin{figure}[t]
\centering
$
\xymatrix@C=3pc@R=.8pc{
\tau
\ar^{\hspace*{-1em}s}[r]
\ar_{\coiter\;s}[d]    &
\tau\;\ty{D}
\ar^{\map_\ty{D}\;(\coiter\;s)}[d]  \\
\langT{\alpha}
\ar^{\hspace*{-1em} \langle \finalOp,\, \derivOp\rangle}[r]  &
\langT{\alpha}\;\ty{D}
}
$
\caption{Unique morphism $\coiter\;s$ to the
  final coalgebra $(\langT{\alpha},\,\langle \finalOp,\, \derivOp\rangle)$}
\label{fig-coiter}
\end{figure}

\begin{figure}[t]
    \centering\vspace{-.5ex}
    $
    \xymatrix@C=3pc@R=.8pc{
        \tau
        \ar^{\hspace*{-1em}s}[r]
        \ar_{\corec\;s}[d]    &
        (\langT{\alpha} + \tau)\;\ty{D}
        \ar^{\map_\ty{D}\;[\Id,\,\corec\;s]}[d]  \\
        \langT{\alpha}
        \ar^{\hspace*{-1em} \langle \finalOp,\, \derivOp\rangle}[r]  &
        \langT{\alpha}\;\ty{D}
    }
    $
    \caption{Characteristic theorem of the corecursor}
    \label{fig-corec}
\end{figure}

Corecursion differs from coiteration by additionally allowing the user to stop the
coiteration process by providing a fixed non-corecursive value. In Isabelle this
is mirrored by another combinator: the \emph{corecursor} $\corec \COLON (\tau
\to (\langT{\alpha}+\tau)\;\ty{D}) \to \tau \to \langT{\alpha}$ where the sum
type $+$ offers the possibility either to continue corecursively as before
(represented by the type $\tau$) or to stop with a fixed value of type
$\langT{\alpha}$. The corecursor satisfies the characteristic property shown in
Figure~\ref{fig-corec}, where the square brackets denote a case distinction on
$+$, i.e.\ $[f,\,g]\;x = \Case\;x\;\Of\;\Inl\;l \Rightarrow f\;l \mid \Inr\;r
\Rightarrow g\;r$.
Corecursion is not more expressive than coiteration (since $\corec$ can be
defined in terms of $\coiter$), but it is more convenient to use. For
instance, the non-corecursive specifications of $\One$ and $\AtomOp$, and the
$\Else$ branch of $\TimesLROp$ exploit this additional flexibility.

{ \looseness=-1 The $\Primcorec$ command~\cite{BlanchetteHLPPT-ITP14} reduces a
user specification to a non-recursive definition using the corecursor. For
example, the union operation $\PlusOp$ is internally defined as $\lambda
L~K.~\corec~(\lambda (L,\,K).\;(\final{L} \LAND \final{K},\,\lambda
a.~\const{Inr}\;(\deriv{L}{a},\,\deriv{K}{a})))~(L,\,K)$. The $\ty{D}$-coalgebra
argument to $\corec$ resembles the right hand sides of the selector equations
for $\PlusOp$ (with the corecursive calls replaced by $\Inr$). In fact, for this
simple definition mere coiteration would suffice.
An example that uses the convenience that corecursion provides, is the deferred concatenation $\TimesLROp$. Its internal definition reads: $\lambda
L~K.~\corec~(\lambda (L,\,K).\;(\final{L} \LAND \final{K},\,\lambda (x,\,b).\;\If\;b\;\Then\;\Inr\;(\deriv{L}{x},\,K)\;\Else\;
\If\;\final{L}\;\Then\;\Inr\;(\deriv{K}{x},\,\One)\;\Else\;\Inl\;\Zero)$.
As end users, most of the time we are happy being able to write the high-level
corecursive specifications, without having to explicitly supply coalgebras.

}

It is worth noting that the final coalgebra $\langT{\alpha}$ itself corresponds
to the automaton, whose states are languages, acceptance is given by
$\const{o}\;L = \final{L}$, and the transition function by $\const{d}\;a\;L =
\deriv{L}{a}$.  For these definitions, we obtain $\corec\;\langle \const{o},\,
\const{d} \rangle\;(L : \langT{\alpha}) = L$. For regular languages this
automaton corresponds to the minimal automaton (since equality on tries
corresponds to language equivalence), which is finite by the Myhill--Nerode
theorem. This correspondence is not very practical though, since we typically
label states of automata with something finite, in particular not with languages
(represented by infinite tries).

A second consequence of the finality of $\langT{\alpha}$ is the coinduction
principle that we have seen earlier. It follows from the fact that final
coalgebras are quotients by bisimilarity, where bisimilarity is defined as the
existence of a bisimulation relation.

\section{Discussion and Related Work}
\label{sec:3:discuss}

Our development is a formalized counterpart of Rutten's introduction to the
coalgebraic view on languages~\cite{Rutten98}. In this section we discuss
further related work.

\PARA{Coalgebraic View on Formal Languages}

The coalgebraic approach to languages has recently received some attention.
Landmark results in language theory were rediscovered and generalized. Silva's
recent survey~\cite{silva15} highlights some of those results including the
proofs of correctness of Brzozowski's subtle deterministic finite automaton
minimization algorithm~\cite{BonchiBRS12}. %
The coalgebraic approach yields some algorithmic advantages, too. Bonchi and Pous
present a coinductive algorithm for checking equivalence of non-deterministic
automata that outperforms all previously known algorithms by one order of
magnitude~\cite{BonchiP13}. Another recent development is our formally
verified coalgebraic algorithm for deciding weak monadic second-order logic of
one successor (WS1S)~\cite{Traytel-CSL15}. This formalization employs the
Isabelle library presented here.

\PARA{Formal Languages in Proof Assistants}

The traditional set-of-words view on formal languages is
formalized in most proof assistants. In contrast, we are not aware of any other
formalization of the coalgebraic view on formal languages in a proof assistant.

\looseness=-1 Here, we want to compare our formalization with the Isabelle
incarnation of the set-of-words view developed by Krauss and Nipkow for the
correctness proof of their regular expression equivalence
checker~\cite{KraussN11}. Both libraries are comparably concise. In 500 lines
Krauss and Nipkow prove almost all axioms of Kleene algebra and the
characteristic equations for the left quotients (the $\derivOp$-specifications
in our case). They reuse Isabelle's libraries for sets and lists, which come
with carefully tuned automation setup. Still, their proofs tend to require
additional induction proofs of auxiliary lemmas, especially when reasoning about
iteration. Our formalization is 700 lines long. We prove all axioms of Kleene
algebra and connect our representation to the set-of-words view via the
bijections $\fromLangOp$ and $\toLangOp$. Except for those bijections our
formalization does not rely on any other library. Moreover, when we changed our
5000 lines long formalization of a coalgebraic decision procedure for
WS1S~\cite{Traytel-CSL15} to use the infinite tries instead of the set-of-words
view, our proofs about WS1S became approximately 300 lines shorter. Apparently,
a coalgebraic library is a good fit for a coalgebraic procedure.

Paulson presents a concise formalization of automata theory based
on hereditarily finite sets~\cite{Paulson15}. For the semantics he reuses Krauss
and Nipkow's set-of-words formalization.

\PARA{Non-Primitive Corecursion in Proof Assistants}

Automation for corecursion in proof assistants is much less developed than its
recursive counterpart. The Coq proof assistant supports corecursion
up to constructors~\cite{Chlipala13}. Looking at our examples, however, this
means that Coq will not be able to prove productivity of the natural
concatenation and iteration specifications automatically, since both go beyond
up-to constructors. Instead, our reduction to primitive corecursion can be
employed to bypass Coq's productivity checker.

Recently, we have added the support for corecursion up to so
called friendly operations to
Isabelle/HOL~\cite{BlanchetteBLPT, BlanchettePT-ICFP15}. (Before this addition, Isabelle supported only primitive corecursion~\cite{BlanchetteHLPPT-ITP14}.) An operation is friendly if, under lazy
evaluation, it does not peek too deeply into its arguments, before producing at
least one constructor. %
For example, the friendly operation $\Plus{L}{K} = \LangC\;(\final{L} \LOR
\final{K})\;(\lambda x.\;\Plus{\deriv{L}{x}}{\deriv{K}{x}})$ destructs only one
layer of constructors, in order to produce the topmost $\LangC$.
In contrast, the primitively corecursive
equation $\mathsf{deep}\;L = \LangC\;(\final{L})\;(\lambda
x.\;\mathsf{deep}\;(\deriv{(\deriv{L}{x})}{x}))$ destructs two layers of
constructors (via $\derivOp$) before producing one and is therefore not
friendly. Indeed, we will not be able to reduce the equation $\mathsf{bad} =
\LangC\;\True\;(\lambda\_.\;\mathsf{deep}\;\mathsf{bad})$ (which is corecursive
up to $\mathsf{deep}$) to a primitively corecursive specification. And there is
a reason for it: $\mathsf{bad}$ is not uniquely specified by the above equation,
or in other words not productive.

Since $\PlusOp$
is friendly, and $\TimesOp$ and $\ShuffleOp$ are corecursive up to $\PlusOp$, this new infrastructure
allows us to use the constructor view version of the natural selector equations~\eqconc{} and \eqshuffle{} for $\TimesOp$ and $\ShuffleOp$ instead of the more complicated primitively corecursive definitions from
Section~\ref{sec:3:corec}.
\begin{quotex}
$\keyw{corec}~(\keyw{friend})~\PlusOp {}\COLON \langT{\alpha}\to\langT{\alpha}\to\langT{\alpha}~\Where\\
\TAB\Plus{L}{K} = \LangC\;(\final{L} \LOR \final{K})\;
(\lambda x.\;\Plus{\deriv{L}{x}}{\deriv{K}{x}})$\\[1\jot]
$\keyw{corec}~(\keyw{friend})~\TimesOp {}\COLON \langT{\alpha}\to\langT{\alpha}\to\langT{\alpha}~\Where\\
\TAB\Times{L}{K} = \LangC\;(\final{L} \LAND \final{K})\;
(\lambda x.\;\Plus{(\Times{\deriv{L}{x}}{K})}{
    (\If\;\final{L}\;\Then\;\deriv{K}{x}\;\Else\;\Zero)})$\\[1\jot]
$\keyw{corec}~(\keyw{friend})~\ShuffleOp {}\COLON \langT{\alpha}\to\langT{\alpha}\to\langT{\alpha}~\Where\\
\TAB\Shuffle{L}{K} = \LangC\;(\final{L} \LAND \final{K})\;
(\lambda x.\;\Plus{(\Shuffle{\deriv{L}{x}}{K})}{(\Shuffle{L}{\deriv{K}{x}})})$
\end{quotex}
The $\keyw{corec}$ command defines the specified constants and the $\keyw{friend}$ option registers them as friendly operations by automatically discharging the emerging proof obligations ensuring that the operations
consume at most one constructor to produce one constructor. Since $\TimesOp$ is friendly, too, we can define the corecursive up to $\TimesOp$ iteration $\StarOp$ using its natural equations~\eqstar.
\begin{quotex}
$\keyw{corec}~(\keyw{friend})~\Star{\_} {}\COLON \langT{\alpha}\to\langT{\alpha}~\Where\\
\TAB\Star{L} = \LangC\;\True\;
(\lambda x.\;\Times{\deriv{L}{x}}{\Star{L}})$
\end{quotex}

Internally, \keyw{corec} reduces the corecursive specification to a primitively corecursive one following an abstract, category theory inspired construction. Yet, what
this abstract construction yields in practice is relatively close to our manual construction for
concatenation. (In contrast, the iteration case takes some shortcuts, which the
abstract view does not offer.)

{
\looseness=-1
On the reasoning side, \keyw{corec} provides some automation, too. It automatically derives the corresponding coinduction up-to rules for the registered (sets of) friendly operations.
Overall, the usage of \keyw{corec} compresses our development
from 700 to 550 lines of Isabelle text.

}

{\setlooseness{-1}
Agda's combination of copatterns (i.e., destructor view) and sized
types~\cite{AbelPTS-POPL13,abelP-2013} is the most expressive implemented support for
corecursion in proof assistants to date. However, using sized types often means
that one has to encode proofs of productivity manually in the type of the
defined function. Thus, it is possible to define concatenation and
iteration using their natural equations \eqconc{} and \eqstar{} in Agda. Recently, Abel~\cite{Abel2016CMCS,Abel} has formalized those definitions of regular operations up to proving the
recursion equation $L^* = \Plus{\One}{\Times{L}{\Star{L}}}$ for iteration in 219~lines of Agda text,
which correspond to 125~lines in our version.
His definitions are equally concise as the ones using $\keyw{corec}$, but his proofs require more
manual steps.

}

\section{Conclusion}
\label{sec:concl}

We have presented a particular formal structure for computation and deduction:
infinite tries modeling formal languages. Although this representation is
semantic and infinite, it is suitable for computation---in particular we obtain
a matching algorithm for free on tries constructed by regular operations.
Deduction does not come short either: coinduction is the convenient reasoning
tool for infinite tries. Coinductive proofs are concise, especially for
(in)equational theorems such as the axioms of Kleene algebra.

Codatatypes might be just the right tool for thinking algorithmically about
semantics. We hope to have contributed to their dissemination by outlining some
of their advantages.

\paragraph*{\textbf{Acknowledgments}}

{\relax

I thank Andrei Popescu for introducing me to the coinductive view on formal
languages. I am grateful to Tobias Nipkow and David Basin for encouraging this
work. Jasmin Blanchette, Andrei Popescu, Ralf Sasse, and the anonymous FSCD and LMCS
reviewers gave many precious hints on early drafts of this article helping to
improve both the presentation and the content. Jurriaan Rot suggested important related
work and recommended proving inequalities of tries by coinduction on $\leq$ rather than on $=$.

}

\bibliographystyle{myabbrv}
\bibliography{bib}{}

\end{document}